\documentclass[sigconf,authorversion]{acmart}

\usepackage{booktabs} % For formal tables

\usepackage[utf8]{inputenc}
\usepackage[T1]{fontenc}
\usepackage[position=bottom]{subfig}
\usepackage{balance}
\usepackage{csquotes}
\usepackage[ruled,linesnumbered]{algorithm2e}
\usepackage{listings}
\usepackage[defblank]{paralist}
\usepackage{float}
\usepackage[defblank]{paralist}
\usepackage{tikz}
\usepackage{pgfplots}
\usepackage{xcolor}
\usepackage{colortbl}
\usepackage{multicol}
\usetikzlibrary{positioning,shapes,fit,chains,automata,patterns}

\copyrightyear{2019}
\acmYear{2019}
\setcopyright{iw3c2w3}
\acmConference[WWW '19]{Proceedings of the 2019 World Wide Web Conference}{May 13--17, 2019}{San Francisco, CA, USA}
\acmBooktitle{Proceedings of the 2019 World Wide Web Conference (WWW'19), May 13--17, 2019, San Francisco, CA, USA}
\acmPrice{}
\acmDOI{10.1145/3308558.3313652}
\acmISBN{978-1-4503-6674-8/19/05}

\def\ojoin{\setbox0=\hbox{$\bowtie$}%
  \rule[-.02ex]{.25em}{.4pt}\llap{\rule[\ht0]{.25em}{.4pt}}}
\def\leftouterjoin{\mathbin{\ojoin\mkern-5.8mu\bowtie}}

\newcommand{\sage}{\textsc{SaGe}}
\newcommand{\sagee}{\textsc{SaGe}\xspace}

\begin{document}

\title{\sagee: Web Preemption for  Public SPARQL Query Services}

\author{Thomas Minier}
\orcid{0000-0002-7321-286X}
\affiliation{%
  \institution{LS2N, University of Nantes}
  \city{Nantes}
  \country{France}
}
\email{thomas.minier@univ-nantes.fr}

\author{Hala Skaf-Molli}
\orcid{0000-0003-1062-6659}
\affiliation{%
  \institution{LS2N, University of Nantes}
  \city{Nantes}
  \country{France}
}
\email{hala.skaf@univ-nantes.fr}

\author{Pascal Molli}
\orcid{0000-0001-8048-273X}
\affiliation{%
  \institution{LS2N, University of Nantes}
  \city{Nantes}
  \country{France}
}
\email{pascal.molli@univ-nantes.fr}

\renewcommand{\shortauthors}{T. Minier et al.}

\begin{abstract}
  To provide stable and responsive public SPARQL query services,
  data providers enforce quotas on server usage.  Queries which exceed
  these quotas are interrupted and deliver partial results. Such
  interruption is not an issue if it is possible to resume queries
  execution afterward. Unfortunately, there is no preemption model
  for the Web that allows for suspending and resuming SPARQL queries.
  In this paper, we propose \sagee: a SPARQL query engine based on Web
  preemption.
  \sagee allows SPARQL queries to be suspended by the Web server after
  a fixed time quantum and resumed upon client request.  Web
  preemption is tractable only if its cost in time is negligible
  compared to the time quantum. The challenge is to support the full
  SPARQL query language while keeping the cost of preemption
  negligible.  Experimental results demonstrate that \sagee
  outperforms existing SPARQL query processing approaches by several
  orders of magnitude in term of the average total query execution
  time and the time for first results.
\end{abstract}

\maketitle

\section{Introduction}\label{sec:introduction}

\textbf{Context and motivation:} Following the Linked Open Data
principles (LOD), data providers published billions of RDF
triples \cite{bizer2009linked,schmachtenberg2014adoption}.
However, providing a public service that allows anyone to
execute any SPARQL query at any time is still an open issue. As public
SPARQL query services are exposed to an unpredictable load of
arbitrary SPARQL queries, the challenge is to ensure that the service
remains \emph{available} despite variation in terms of the arrival rate of
queries and \emph{resources} required to process queries.

To overcome this problem, most public LOD providers enforce a fair use
service policy based on \emph{quotas}~\cite{buil2013sparql}. According
to DBpedia administrators:\enquote{A Fair Use Policy is in place in
  order to provide a stable and responsive endpoint for the
  community.}\footnote{\url{http://wiki.dbpedia.org/public-sparql-endpoint}}
The public DBpedia SPARQL endpoint~\footnote{\url{http://dbpedia.org/sparql}} Fair Use
Policy prevents the execution of SPARQL longer than 120 seconds or that return more than
10000 results, with a limit of 50 concurrent connections and 100
requests per second per IP address.
Quotas aim to share fairly server resources among Web clients. Quotas
on communications limit the arrival rate of queries per IP.  Quotas
on space prevent one query to consume all the memory of the server.
Quotas on time aim to avoid the \emph{convoy
  phenomenon}~\cite{BlasgenGMP79}, \emph{i.e.}, a long-running query
will slow down a short-running one, in analogy with a truck on a
single-lane road that creates a convoy of cars.  The main
  drawback of quotas is that \emph{interrupted queries can only deliver
  partial results, as they cannot be resumed}. This is a serious limitation for Linked Data
consumers, that want  to execute long-running
queries~\cite{axeldesemweb18}.

\textbf{Related works:} Existing approaches address this issue by
decomposing SPARQL queries into subqueries that can be executed under
the quotas and produce complete results~\cite{axel2014}.  Finding such
decomposition is hard in the general case, as quotas can be different
from one server to another, both in terms of values and
nature~\cite{axel2014}.  The Linked Data Fragments (LDF)
approach~\cite{olaf2017,verborgh2016triple} tackles this issue by
restricting the SPARQL operators supported by the server.  For
example, in the Triple Pattern Fragments (TPF)
approach~\cite{verborgh2016triple}, a TPF server only evaluates
 triple patterns.  However, LDF approaches generate a large number of
subqueries and substantial data transfer.

\textbf{Approach and Contributions:} We believe that the issue related to
time quotas is not interrupting a query, but the impossibility for the
client to \emph{resume} the query execution afterwards.  In this
paper, we propose \sage, a SPARQL query engine based on Web
preemption. Web preemption is the capacity of a Web server to suspend
a running query after a time quantum with the intention to resume it
later.  When suspended, the state $S_i$ of the query is returned to
the Web client. Then, the client can resume query execution by sending
$S_i$ back to the Web server.

Web preemption adds an overhead for the Web server to suspend the
running query and resume the next waiting query. Consequently, the
main scientific challenge here is to keep this overhead marginal
whatever the running queries, to ensure good query execution
performance.
The contributions of this paper are as follows:
\begin{asparaitem}
\item We define and formalize a \emph{Web preemption} model that allows
  to suspend and resume SPARQL queries.
\item We define a set of preemptable query operators for which we
  bound the complexity of suspending and resuming of these operation, both in time and
  space. This allows to build a preemptive Web server that supports a
  large fragment of the SPARQL query language.
\item We propose \sage, a SPARQL query engine, composed of a
  preemptive Web server and a smart Web client that allows  executing
  full SPARQL queries\footnote{The \sagee software and a demonstration are
    available at \url{http://sage.univ-nantes.fr}}.
\item We compare the performance of the \sagee engine with existing approaches
  used for hosting public SPARQL services. Experimental results
  demonstrate that \sagee outperforms existing
  approaches by several orders of magnitude in term of the average
  total query execution time and the time for first results.
\end{asparaitem}

This paper is organized as follows. Section~\ref{sec:related_work}
summarizes related works.  Section~\ref{sec:approach} defines the Web
preemption execution model and details the \sagee server and the
\sagee client.
Section~\ref{sec:exp_study} presents our experimental results.
Finally, conclusions and future work are outlined in
Section~\ref{sec:conclusion}.

% -------------------------------------------------------
\section{Related Works}\label{sec:related_work}

\paragraph{\textbf{SPARQL endpoints}}

SPARQL endpoints follow the SPARQL
protocol~\footnote{\url{https://www.w3.org/TR/2013/REC-sparql11-protocol-20130321/}}, which
\enquote{describes a means for conveying SPARQL queries and
updates to a SPARQL processing service and returning the results via
HTTP to the entity that requested them}. Without quotas, SPARQL
endpoints execute queries using a First-Come First-Served (FCFS) execution
policy~\cite{Fife68a}. Thus, by design, they can suffer from
\emph{convoy effect}~\cite{BlasgenGMP79}: one long-running query
occupies the server resources and prevents  other queries from executing,
leading to long waiting time and degraded average completion time for
queries.

To prevent convoy effect and ensure a fair sharing of resources among
end-users, most SPARQL endpoints configure quotas on their servers. They mainly
restrict the arrival rate per IP address and limit the execution time  of
queries. Restricting the arrival rate allows end-users to retry
later, however, limiting the execution time leads some queries to deliver only partial
results. To illustrate, consider the SPARQL query $Q_1$ of
Figure~\ref{fig:motiv_example}.  Without any quota, the total number
of results of $Q_1$ is 35 215, however, when executed against
the DBpedia SPARQL endpoint, we found only 10 000 results out of
35 215~\footnote{All results were obtained on DBpedia version 2016-04.}.

Delivering partial results is a serious limitation for a public SPARQL
service. In \sagee, we deliver complete results whatever the query.
In some way, quotas interrupt queries
without giving the possibility to resume their execution.
\sagee also interrupts queries, but allows data consumers to resume
their execution later on.

\begin{figure}
\begin{lstlisting}[basicstyle=\scriptsize\sffamily,
  language=sparql,numbers=none,frame=none,columns=fixed,
  extendedchars=true,breaklines=true,showstringspaces=false]
PREFIX dbo: <http://dbpedia.org/ontology/>
PREFIX rdfs: <http://www.w3.org/2000/01/rdf-schema#>
SELECT ?actor ?name ?birthPlace WHERE {
  ?actor a dbo:Actor; rdfs:label ?name; dbo:birthPlace ?city.
  ?city a dbo:City; rdfs:label ?birthPlace.
}
\end{lstlisting}
  \caption{SPARQL Query $Q_1$: finds all actors' birth cities.}
  \label{fig:motiv_example}
\end{figure}

\paragraph{\textbf{Decomposing queries and restricting server interfaces}}

Evaluation strategies~\cite{axel2014} have been studied
for federated SPARQL queries evaluated under quotas. Queries are decomposed into a set of
subqueries that can be fully executed under quotas. The main drawbacks
of these strategies are:
\begin{inparaenum}[(i)]
  \item They need to know which quotas are configured.
  Knowing all quotas that a data provider can implement is not always
  possible.
  \item They can only be applied to a specific class of
  SPARQL queries, \emph{strongly bounded SPARQL queries}, to ensure
  complete and correct evaluation results.
\end{inparaenum}

The Linked Data Fragments (LDF)~\cite{verborgh2016triple,olaf2017}
restrict the server interface to a fragment of the SPARQL
algebra, to reduce the complexity of queries evaluated by the server.
LDF servers are no more compliant with the W3C SPARQL protocol, and SPARQL query processing is distributed between smart clients and
LDF servers. Hartig et al.~\cite{olaf2017} formalized this approach using
Linked Data Fragment machines (LDFMs).
The Triple Pattern Fragments (TPF) approach~\cite{verborgh2016triple}
is one implementation of LDF where the
server only evaluates \emph{paginated triple pattern queries}. As
paginated triple pattern queries can be evaluated in bounded
time~\cite{HelingAMS18}, the server does not suffer from the convoy
effect.  However, as joins are performed on the client, the intensive
transfer of intermediate results leads to poor SPARQL query execution
performance.  For example, the evaluation of the query $Q_1$, of
Figure~\ref{fig:motiv_example}, using the TPF approach generates $507 156$ subqueries
and transfers 2Gb of intermediate results in more than
2 hours.  The Bindings-Restricted Triple Pattern Fragments (BrTPF)
approach \cite{hartig2016bindings} improves the TPF approach by using
the bind-join algorithm \cite{HaasKWY97Optimizing}
to reduce transferred data but joins still executed by the client.  In this paper, we
explore how Web preemption  allows the server to execute a larger
fragment of the SPARQL algebra, including joins,
without generating convoy effects. Processing joins on server side allow to drastically
reduce transferred data  between client and server and improve
significantly performance. For example, \sagee executes the query
$Q_1$ of Figure~\ref{fig:motiv_example} in less than 53s,
with 553 requests and $2.1$Mb transferred.

\paragraph{\textbf{Preemption and Web preemption}}
FCFS scheduling policies and the convoy
effect~\cite{BlasgenGMP79} have been heavily studied in operating
systems. In a system where the duration of tasks vary, a long-running
task can block all other tasks, deteriorating the average completion
time for all tasks. The \emph{Round-Robin} (RR)
algorithm~\cite{kleinrock1964analysis} provides a fair allocation of
CPU between tasks, avoids convoy effect, reduces the waiting time and
provides good responsiveness.  RR runs a task for a given \emph{time
  quantum}, then suspends it and switches to the next task. It
repeatedly does so until all tasks are finished. The value of this
time quantum is critical for performance: when too high, RR behaves
like FCFS with the same issues, and when its too low, the overhead of
context switching dominates the overall performance.  The action of
suspending a task with the intention of resuming it later is called
preemption.
In public Web servers, preemption is already provided by the operating
systems, but \emph{only between running tasks}, excluding those waiting
in the server's queue. If we want to build a fully \emph{preemptive
  Web server}, we need to consider the queries sent to the server as
the tasks and the Web server as the resource that tasks competed to
get access to. In this paper, we explore how preemption can be defined
at the Web level to execute SPARQL queries.

% -------------------------------------------------------
\section{Web preemption Approach}\label{sec:approach}

\begin{figure}
  \centering
     \begin{tikzpicture}[scale=0.1]
  	\draw (10, 5) node {\textbf{FCFS}};
    % draw horizontal line   
    \draw (0,0) -- (70,0);

    % draw vertical lines
    \foreach \x in {0,60,65,70} {
      \draw (\x cm, 0pt) node{$ | $} ;
      \draw (\x cm,3pt) node[below=3pt] { \small{\x}} ;
    }

    % draw nodes
    \draw (30,0) node[above=3pt] {$Q_1$};
    \draw (62.5,0) node[above=3pt] {$Q_2$};
    \draw (67.5,0) node[above=3pt] {$Q_3$};
  \end{tikzpicture}
     \begin{tikzpicture}[scale=0.1]
 	\draw (10, -10) node {\textbf{Web preemption}};
    % draw horizontal line
    \draw (0,0) -- (76,0);

    % draw vertical lines
     \foreach \x in {30,43}{
       \filldraw[pattern=north east lines] (\x,0) rectangle (\x+3,3);
      }
    \foreach \x in {0,30,33,38,43,46,76} {
      \draw (\x cm, 0pt) node{$ | $} ;
      \draw (\x cm,3pt) node[below=3pt] {\small{\x}} ;
    }

    % draw nodes
    \draw (15,0) node[above=1pt] {$ Q_1  $};
    \draw (35.5,0) node[above=1pt] (q2) {$ Q_2 $};
    \draw (40.5,0) node[above=1pt] {$ Q_3 $};
    \draw (61,0) node[above=1pt] {$ Q_1 $};
 \end{tikzpicture}
  \caption{First-Come First-Served (FCFS) policy compared to Web Preemption (time quantum of 30s and overhead of 3s).}
  \label{fig:query_automata}
    \label{fig:fcfsvsrr}
\end{figure}

We define \emph{Web preemption} as the capacity of a Web server to
suspend a running query after a fixed quantum of time and resume the
next waiting query. When suspended, partial results and the state of the suspended query
$S_i$ are returned to the Web client~\footnote{$S_i$ can be returned
  to the client or saved server-side and returned by reference.}. The client can resume query execution by
sending $S_i$ back to the Web server. Compared to a
First-Come First-Served (FCFS) scheduling policy, Web preemption
provides \emph{a fair allocation of Web server resources across
  queries, a better average query completion time per query and a
  better time for first results} \cite{RR2014}. To illustrate,
consider three SPARQL queries $Q_1, Q_2,$ and $Q_3$ submitted
concurrently by three different Web clients. $Q_1, Q_2, Q_3$ execution
times are respectively 60s, 5s and 5s. Figure~\ref{fig:fcfsvsrr}
presents a possible execution of these queries with a FCFS policy on
the first line.  In this case, the throughput of FCFS is
$\frac{3}{70}=0.042$ queries per second, the average completion time
per query is $\frac{60+65+70}{3}=65s$ and the average time for first
results is also 65s.
The second line describes the execution of $Q_1-Q_3$ using Web
preemption, with a time quantum of 30s. We consider a preemption
overhead of 3s (10\% of the quantum).  In this case, the throughput is
$\frac{3}{76}=0.039$ query per second but the average completion time
per query is $\frac{76+38+43}{3}=52.3s$ and the average time for first results
is approximately $\frac{30+38+43}{3}=37s$. If the quantum is set to
60s, then Web preemption is equivalent to FCFS. If the quantum is too
low, then the throughput and the average completion time are
deteriorated.

Consequently, the challenges with Web preemption are \emph{to bound the preemption overhead in time and space} and \emph{determine the time quantum} to amortize the overhead.

\subsection{Web Preemption Model}
\label{sec:web-preemtion-model}

\begin{figure}
  \centering
  \resizebox{0.4\textwidth}{!}{%
    \begin{tikzpicture}[shorten >=1pt, node distance=3cm, on grid,every initial by arrow/.style={-latex,thick}]
  \tikzstyle{smaller}=[inner sep=0.5pt, minimum size=20pt, ellipse, align=center]
  \tikzstyle{exec_path}=[thick, draw=blue]

	\node[initial, state, smaller] (new) {New\\query};
	\node[state, smaller, above=2cm of new] (waiting) {Waiting\\in queue};
	\node[state, smaller, above=2cm of waiting] (running) {Running};
	\node[state, smaller, right=of waiting] (save) {Suspended};
	\node[state, smaller, below=2cm of save] (suspend) {Saved};
	\node[accepting, state, smaller, right=of suspend] (done) {Done};

	\node[draw,thick,dotted,fit=(waiting) (running) (save),label=above:{\textbf{Preemptive Web Server}}] {};
	\node[draw,thick,dotted,fit=(new) (suspend) (done),label=below:{\textbf{Smart Web Client}}] {};

	\path[->,-latex,thick]
		(new) edge node [left, align=center, text width=1.5cm] {$Q$ sent to server} (waiting)
		(waiting) edge node [left=1mm, align=center, text width=1.2cm, fill=white] {Worker available} (running)
		(running) edge [bend left] node [above, sloped, align=center, text width=2cm, fill=white] {Execution completed} (done)
		(running) edge node [below, sloped, align=center, text width=2cm] {Quantum exhausted} (save)
		(save) edge node [right=1mm, align=center, text width=1.5cm] {Server sends $S_i$} (suspend)
		(suspend) edge node [below=1mm, sloped, align=center, text width=2cm, fill=white] {Client sends $S_i$} (waiting);
\end{tikzpicture}
  }
  \caption{Possible states of a query execution in a preemptive Web Server.}
  \label{fig:query_automata}
\end{figure}
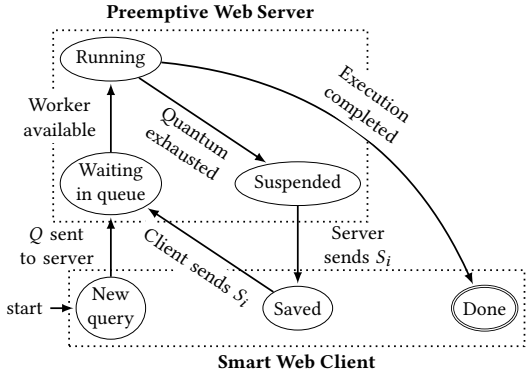

We consider a \emph{preemptive Web server}, hosting RDF datasets, and
a \emph{smart Web client}, that evaluates SPARQL queries using the
server.
For the sake of simplicity, we only consider
\emph{read-only queries}~\footnote{Preemption and concurrent updates
  raise issues on correctness of results.} in this paper.  The server has a pool of
\emph{server workers} parameterized with a fixed time quantum. Workers are  in charge of
queries execution. The server has also a \emph{query queue} to store
incoming queries when all workers are busy. We consider an
\emph{infinite} population of clients, a \emph{finite} server queue
and a \emph{finite} number of Web workers.

The preemptive Web server suspends and resumes queries after the time quantum. A running query $Q_i$ is represented by its \emph{physical query
  execution plan}, denoted $P_{Q_i}$. Suspending the execution of $Q_i$ is an operation applied on
$P_{Q_i}$ that produces a \emph{saved state} $S_i$;
$\texttt{Suspend}(P_{Q_i}) = S_i$.  Resuming the execution of a query
$Q_i$ is the inverse operation, it takes $S_i$ as parameter and restores the
physical query execution plan in its
suspended state. Therefore, the preemption is correct if
$\texttt{Resume}(\texttt{Suspend}(P_{Q_i})) = P_{Q_i}$.

Figure~\ref{fig:query_automata} presents  possible states of a
query. The transitions are executed either by the Web server or by the
client.

The Web server accepts, in its waiting queue, Web requests containing
either SPARQL queries $Q_i$, or suspended queries $S_i$. If a worker
is available, it picks a query in the waiting queue. For  $Q_i$,
the worker produces a physical query execution plan $P_{Q_i}$ using
the \emph{optimize-then-execute} \cite{graefe1993query} paradigm and
starts its execution for the time quantum.  For  $S_i$, the
server resumes the execution of $Q_i$.  The time to produce or resume
the physical query execution plan for a query is not deducted from the
quantum.

If a query terminates before the time quantum, then results are
returned to the Web client. If the time quantum is exhausted and the
query is still running, then the \texttt{Suspend} operation is
triggered, producing a state $S_i$ which is returned to the Web client
with partial results. The time to suspend the query is not deducted
from the quantum.  Finally, the Web client is free to continue the
query execution by sending $S_i$ back to the Web server.

The main source of overhead in this model is the time and space complexity of
the \texttt{Suspend} and \texttt{Resume} operations,  \emph{i.e.},  time to stop
and save the running query followed by the time to resume the next
waiting query.
 Our objective is to bound these complexities such that they depend only on  the \emph{query
  complexity}, \emph{i.e.}, the number of operations in the query plan.
  Consequently, \emph{the problem is to
  determine which physical query plans $P_{Q_i}$ have a preemption
  overhead bounded in $\mathcal{O}(|Q|)$}, where $|Q|$ denotes the
number of operators in the expression tree of $P_{Q_i}$.

\subsection{Suspending and Resuming Physical Query Execution Plans}\label{sec:preemption_sparql}

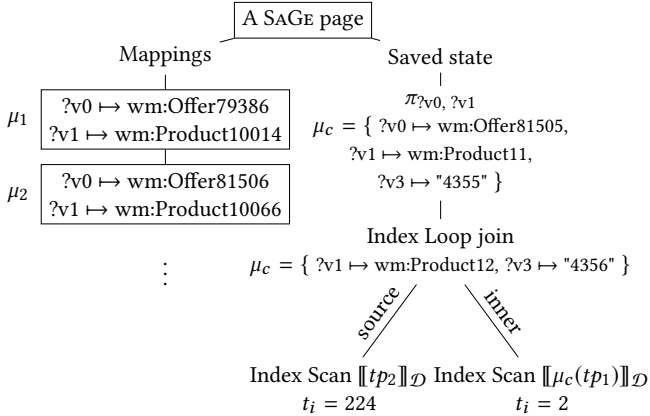
\begin{figure}
  \centering
  \subfloat [SPARQL query $Q_2$, from the Waterloo SPARQL Diversity Test suite benchmark~\cite{alucc2014diversified}] {\label{query:q2}
  \resizebox{0.45\textwidth}{!}{%
\begin{lstlisting}[
       basicstyle=\scriptsize\sffamily,
       language=sparql,
       extendedchars=true
    ]
PREFIX schema: <http://schema.org/contentSize/>
PREFIX gr: <http://purl.org/goodrelations/>
SELECT DISTINCT ?vo ?v1 WHERE {
 ?v0 gr:includes ?v1. # tp1
 ?v1 schema:contentSize ?v3. # tp2
}
\end{lstlisting}
  }
  }\vfill
  \subfloat [One page returned by the \sagee server during $Q_2$ evaluation.] {\label{fig:sage_page}
    \resizebox{0.5\textwidth}{!}{%
      \begin{tikzpicture}
  \tikzstyle{level 1}=[level distance=0.5cm,sibling distance=3.8cm]
  \tikzstyle{level 2}=[level distance=1.2cm, sibling distance=4.5cm]
  \tikzstyle{level 3}=[level distance=1.5cm, sibling distance=4.5cm]
  \tikzstyle{level 4}=[level distance=1.9cm, sibling distance=2.8cm]
  \node [draw] (root) {A \sagee page}
  child { node (bindings) {Mappings}
  }
  child {
  	node {Saved state}
    	child {
          node [align=center] {$\pi_{\text{?v0, ?v1}}$ \\
            $\mu_c = \{$ {\small ?v0 $\mapsto$ wm:Offer81505,} \\
          {\small ?v1 $\mapsto$ wm:Product11,} \\
          {\small ?v3 $\mapsto$ "4355"} $\}$ }
		child { node [align=center] {Index Loop join \\ $\mu_c = \{$ {\small ?v1 $\mapsto$ wm:Product12,} {\small ?v3 $\mapsto$ "4356"} $\}$ }
      		  child { node [align=center]{Index Scan $\llbracket tp_2 \rrbracket_\mathcal{D}$ \\$t_i = 224$}
        edge from parent node [midway,sloped, above] {source}
            }
            child { node [align=center]{Index Scan $\llbracket \mu_c(tp_1) \rrbracket_\mathcal{D}$ \\$t_i = 2$}
              edge from parent node [midway,sloped, above] {inner}
            }
          }
	}
};

  \node [draw, align=center, below=2mm of bindings] (b1) {
      ?v0 $\mapsto$ wm:Offer79386 \\
      ?v1 $\mapsto$ wm:Product10014
  };

  \node [draw, align=center, below=2mm of b1] (b2) {
      ?v0 $\mapsto$ wm:Offer81506 \\
      ?v1 $\mapsto$ wm:Product10066
  };

  \node [left=0.5mm of b1] {$\mu_1$};
  \node [left=0.5mm of b2] {$\mu_2$};

  \path [draw]
    (bindings) edge (b1)
    (b1) edge (b2);

  \node [align=center, below=2mm of b2] (bdots) {$\vdots$};

\end{tikzpicture}
    }
  }
  \caption{A tree representation of a page returned by the \sagee
    server when executing SPARQL query $Q_2$ with the saved plan
    passed by value.}
  \label{fig:q1_page}
\end{figure}

The \texttt{Suspend} operation is applied to a running physical query
execution plan $P_{Q_i}$. It is obtained from the logical query
execution plan of $Q_i$ by selecting physical operators that implement
operators in the logical plan~\cite{garcia2008database}. A running $P_{Q_i}$ can be represented as an expression tree of physical
operators where each physical operator has a state, \emph{i.e.}, all
internal data structures allocated by the operator.
Suspending a plan $P_{Q_i}$ requires to traverse the expression tree
of the plan and save the state of each physical operator. To illustrate,
consider the execution of query $Q_2$ in Figure \ref{query:q2}. The state of
$P_{Q_2}$ can be saved as described in Figure~\ref{fig:sage_page}: the
state of the Scan operator is represented by the $id$ of the last
triple read $t_i$, the state of the Index Loop Join operator is
represented by mappings pulled from the previous operator and the
state of the inner scan of $tp_1$.
Suspending and resuming a physical query execution plan raise several
major issues.

\paragraph{\textbf{Suspending physical operators in constant time}}

Bounding the time complexity of $Suspend(P_{Q_i})$ to $O(|Q_i|)$
requires to save the state of all physical operators in constant
time. However, \emph{some physical operators need to materialize data},
\emph{i.e.}, build collections of mappings collected from other operators
to perform their actions. We call these operators
\emph{full-mappings} operators, in opposition to
\emph{mapping-at-a-time} operators that only need to consume one
mapping at a time from child operators\footnote{This is a clear
reference to \emph{full-relation} and \emph{tuple-at-a-time}
operators in database systems~\cite{garcia2008database}[Chapter 15.2]}.
Saving the state of full-mappings operators cannot be done in
constant time. To overcome this problem, we distribute operators between a \sagee server
and a \sagee smart client as follows:

\begin{compactitem}

\item \emph{Mapping-at-a-time} operators are suitable for Web preemption, so they are supported by the \sagee server. These operators are a subset of \textsc{CoreSPARQL}~\cite{olaf2017}, composed of Triple Patterns, AND, UNION, FILTER and SELECT.
  We explain how to implement these operators server-side in Section~\ref{sec:sagee-preempt-serv}.

\item \emph{Full-mappings} operators do not support Web preemption,
  so they are implemented in the \sagee smart client. These operators are:
  OPTIONAL, SERVICE, ORDER BY, GROUP BY, DISTINCT, MINUS, FILTER
  EXIST and aggregations (COUNT, AVG, SUM, MIN, MAX). We explain
  how to implement these operators in the smart client in Section~\ref{sec:sagee-smart-web}.
\end{compactitem}

As proposed in LDF~\cite{olaf2017,verborgh2016triple}, the
collaboration of the \sagee smart client and the \sagee server allows
to support full SPARQL queries.

\paragraph{\textbf{Communication between operators}}
Bounding the time complexity of $Suspend(P_{Q_i})$ to $O(|Q_i|)$
requires to avoid materialization of intermediate results when
physical operators communicate. This only concerns operators of the
\sagee server.
To solve this issue, we follow \emph{the iterator
  model}~\cite{GraefeM93,graefe1993query}. In this model, operators
are connected in a \emph{pipeline}, where they are chained together in
a \emph{pull-fashion}, such as one iterator pulls solution mappings
from its predecessor(s) to produce results.

\paragraph{\textbf{Saving consistent states of the physical query plan}}
Some physical query operators have critical sections and cannot be
interrupted until exiting those sections. Saving the physical plan
in an inconsistent state leads to incorrect preemption. In other words,
we could have $Resume(Suspend(P_{Q_i})) \neq P_{Q_i}$.
To solve this issue, we have to detail, for each  physical
operator supported by the \sagee server, where are located critical sections and
estimate if the waiting time for exiting the critical section is
acceptable.

\paragraph{\textbf{Resuming physical operators in constant time}}
Reading a saved plan as the one presented in
Figure~\ref{fig:sage_page} should be in $O(|Q|)$. The \emph{Index Scan
  $\llbracket tp_2 \rrbracket_\mathcal{D}$} has been suspended after
reading the triple with $id=224$. Resuming the scan requires that the
access to the triple with $id \geq 224$ is in constant time. This can
be achieved with adequate indexes.
To solve this issue, we have to define clearly for each physical
operator what are the requirements on backend to bound the overhead of
resuming.

\subsection{The \sagee Preemptable Server}
\label{sec:sagee-preempt-serv}

The \sagee server supports  Triple Patterns,
AND, UNION, FILTER and SELECT operators. The logical
and physical query plans are builds thanks to the
\emph{optimize-then-execute} \cite{graefe1993query} paradigm. The main
issues here are the cost of the \texttt{Suspend} and \texttt{Resume} operations
on the physical plan and how to interrupt physical operators.

To support preemption, we extend classical iterators to
\emph{preemptable iterators}, formalized in
Definition~\ref{def:preemptable_operator}. As iterators are connected in a pipeline,
we consider that each iterator is also responsible for recursively
stopping, saving and resuming its predecessors.

\begin{definition}[Preemptable iterator]\label{def:preemptable_operator}
  A \emph{preemptable iterator} is an iterator that supports, in addition to the classic \texttt{Open}, \texttt{GetNext} and \texttt{Close} methods \cite{GraefeM93}, the following methods:
  \begin{compactitem}
  	\item \texttt{Stop}: interrupts the iterator and its
          predecessor(s). \texttt{Stop} waits for all non interruptible sections to complete.
  	\item \texttt{Save}: serializes the current state of the iterator and its predecessor(s) to produce a \emph{saved state}.
  	\item \texttt{Load}: reloads the iterator and its predecessor(s)
          from a saved state.
  \end{compactitem}
\end{definition}

\begin{algorithm}[t]
  \SetAlgoVlined
  \SetKwInput{Input}{Require}
  \SetKwBlock{Synchro}{synchronized}{{end synchronized}}
  \SetKwProg{Fn}{Function}{\string:}{}
  \SetKwComment{tcp}{}{}%
  \SetKwComment{tcc}{// }{}%

  \DontPrintSemicolon
  \SetInd{0.55em}{0.55em}
  \caption{Implementation of the \texttt{Suspend} and \texttt{Resume} functions following the iterator model}
  \label{algo:plan_preemption}

  \Input{
    $\mathcal{I}$: pipeline of iterators,
    $S$: serialized pipeline state (as generated by \texttt{Suspend})
  }
  \setlength\columnsep{0pt}
  \vspace{-0.4cm}
\begin{multicols}{2}
  \Fn{\texttt{Suspend}($\mathcal{I}$)}{
    \textbf{let} root$\leftarrow$first iterator in $\mathcal{I}$\;
    \textbf{Call} root.Stop() \;
    \Return{root.Save()}
  }

  \Fn{\texttt{Resume}($\mathcal{I}, S$)}{
    \textbf{let} root$\leftarrow$first iterator in $\mathcal{I}$\;
    \textbf{Call} root.Load($S$) \;
    \Return{$\mathcal{I}$} \;
  }
  \end{multicols}   \vspace{-0.2cm}

\end{algorithm}
 Algorithm~\ref{algo:plan_preemption}
presents the implementation of the \texttt{Suspend} and
\texttt{Resume} functions for a pipeline of preemptable query
iterators.  \texttt{Suspend} simply stops and saves recursively each
iterator in the pipeline and \texttt{Resume} reloads the pipeline in
the suspended state using a serialized state.  To illustrate, consider
the \sagee page of Figure~\ref{fig:sage_page}. This page contains the  plan that evaluates the
SPARQL query $Q_2$  of size $|Q_2|=4$, using \emph{index loop
  joins}~\cite{graefe1993query}.   The
evaluation of $tp_2$ has been preempted after scanning 224 solution
mappings. The index loop join with $tp_1$ has been preempted after
scanning two triples from the inner loop.

In the following, we review SPARQL operators implemented by the
\sagee server as preemptable query iterators.   We modify the
regular implementations of these operators to include non
interruptible section when needed.  Operations left unchanged are not
detailed.  Table~\ref{tab:complex} resumes the complexities related to
the preemption of server operators, where  $\emph{var }(P)$ denotes the set of variables in $P$ as defined in~\cite{perez2009semantics}.

\begin{table}
  \centering
  \resizebox{0.48\textwidth}{!}{
\begin{tabular}{|p{1.7cm}|c|c|p{2.1cm}|} \hline
  \textbf{Preemptable}    & \textbf{Space complexity} & \textbf{Time complexity of} & \textbf{Remarks}  \\
  \textbf{iterator}           & \textbf{of local state} & \textbf{loading local state} &  \\  \hline
  %------------------------------------------------
  $\pi_{v_1, \dots, v_k}(P)$         & $\mathcal{O}(k + |\emph{var }(P)|)$      & $\mathcal{O}(1)$  & \\ \hline
  %------------------------------------------------
 Index Scan $tp$  &  $\mathcal{O}(|tp| + |id|)$       & $\mathcal{O}(\log_b(|\mathcal{D}|))$
                                       & Require indexes on all kinds
                                         of triple patterns \\ \hline
  %------------------------------------------------
  Merge Join $P_1 \bowtie P_2$ & $\mathcal{O}(|\emph{var }(P_1)| + |\emph{var}(P_2)|)$    & $\mathcal{O}(1)$  & \\ \hline
  %------------------------------------------------
  Index Loop Join $P \bowtie tp$ & $\mathcal{O}(|\emph{var }(P)| + |tp| + |id|)$  & $\mathcal{O}(\log_b(|\mathcal{D}|))$  & \\ \hline
  %------------------------------------------------
  $P_1$ UNION $P_2$ & $\mathcal{O}(1)$  & $\mathcal{O}(1)$  & Multi-set Union \\ \hline
  %------------------------------------------------
  $P$ Filter $\mathcal{R}$ & $\mathcal{O}(|\emph{var }(P)| + |\mathcal{R}|)$  &  $\mathcal{O}(1)$  & Pure logical expression only \\
  \hline
  \hline
  Server physical plan & $\mathcal{O}(|Q| \times \log_2(|\mathcal{D}|)$  &  $\mathcal{O}(|Q|\times\log_b(|\mathcal{D}|))$ &  \\
  \hline
  \end{tabular}
  }
  \caption{Complexities of preemption of physical query
    iterators. $|id|$ and $|tp|$ denote the size of encoding an index
    key and a triple pattern, respectively.}
\label{tab:complex}
\end{table}

\begin{algorithm}
  \SetAlgoVlined
  \SetKwInput{Require}{Require}
  \SetKwInput{Data}{Data}
  \SetKwBlock{Synchro}{non interruptible}{{end synchronized}}
  \SetKwProg{Fn}{Function}{\string:}{}
  \SetKwComment{tcp}{}{}%
  \SetKwComment{tcc}{// }{}%

  \DontPrintSemicolon
  \SetInd{0.55em}{0.55em}
  \caption{A \textit{Preemptable Select Iterator}}
  \label{algo:proj}

  \Require{
    $V = \{ v_1, \dots, v_k \}$: projection variables,
    $\mathcal{I}$: predecessor in the pipeline
  }

  \Data{
  $\mu \leftarrow nil$
}
 \vspace{-0.5cm}
\begin{multicols}{2}
  \Fn{GetNext()}{
  	\If{$\mu = nil$}{
  		$\mu \leftarrow$ $\mathcal{I}$.GetNext() \;
	}
  	\Synchro{\label{line:proj_start}
		\textbf{let} $\mu'\leftarrow$ Proj($\mu, V$) \;
		$\mu \leftarrow nil$ \;
		\Return $\mu'$ \label{line:proj_end}
  	}
  }

  \Fn{Save()}{
    \Return $\mu$ \;
  }

  \Fn{Load($\mu'$)}{
  	$\mu \leftarrow \mu'$ \;
      }
   \end{multicols}      \vspace{-0.2cm}

\end{algorithm}

\paragraph{\textbf{SELECT operator}:} The projection operator
$\pi_V(P)$~\cite{Schmidt0L10}, also called SELECT, performs the
projection of mappings obtained from its predecessor $P$ according to
a set of projection variables $V = \{ v_1, \dots, v_k \}$.
Algorithm~\ref{algo:proj} gives the implementation of this
iterator.
In this algorithm,  Lines \ref{line:proj_start}-\ref{line:proj_end} are non-interruptible. This is necessary to
ensure that the operator does not discard mappings without applying
projection to them. Consequently, after the termination of $Suspend(P)$,
$\pi_V(P)$ can be suspended in $\mathcal{O}(k + |\emph{var}(P)|)$ in
space. The projection iterator reloads its local state in $O(1)$.

\paragraph{\textbf{Triple Pattern:}}
The evaluation of a triple pattern $tp$ over $D$
$\llbracket tp \rrbracket_\mathcal{D}$\ \cite{perez2009semantics} is the set of solution mappings of $tp$.
$\mu$ is a solution mapping if  $\mu(tp) = t$ where
 $t$ is the triple obtained by replacing the variables in $tp$ according to $\mu$, such that $t \in \mathcal{D}$.
 The evaluation of $\llbracket tp \rrbracket_\mathcal{D}$ requires to read $\mathcal{D}$ and produce the corresponding set of solutions mappings.

The triple pattern operator supposes that data are accessed using
\emph{index scans} and a clustered index \cite{garcia2008database}.
Algorithm~\ref{algo:index_scan} gives the implementation of a
preemptable index scan for evaluating a triple pattern.  We omitted
the \texttt{Stop()} operation, as the iterator does not need to do any
action to stop.  The \texttt{Save()} operation stores the
triple pattern $tp$ and the index's $id(t)$ of the last triple $t$ read.
Thus, suspending an index scan is done in constant time and the size of the operator state
is in $\mathcal{O}(|tp| + |id(t)|)$.

The \texttt{Load()} operation uses the saved index's $id(t)$ on $tp$ to
find the last matching RDF triple read.  The time complexity of this
operation predominates the time complexity of \texttt{Resume}. This complexity depends on the type of index used. With a
B+-tree \cite{Comer79}, the time complexity of resuming is bound to
$\mathcal{O}(\log_b(|\mathcal{D}|))$.  B+-tree indexes are offered by
many RDF data storage systems
\cite{ErlingM09,NeumannW10,WeissKB08}.

\begin{algorithm}
  \SetAlgoVlined
  \SetKwInput{Require}{Require}
  \SetKwInput{Data}{Data}
  \SetKwBlock{Synchro}{non interruptible}{{end synchronized}}
  \SetKwProg{Fn}{Function}{\string:}{}
  \SetKwComment{tcp}{}{}%
  \SetKwComment{tcc}{// }{}%

  \DontPrintSemicolon
  \SetInd{0.55em}{0.55em}
  \caption{A \textit{Preemptable Index Scan Iterator}, evaluating a triple pattern $tp$ using an index}
  \label{algo:index_scan}

  \Require{
    $tp$: triple pattern,
    $\mathcal{D}$: RDF dataset queried,
    $\mathcal{I}_{tp}$: clustered index over $tp$
  }

  \Data{
  $t$: last matching RDF triple read
  }

  \Fn{GetNext()}{
  	\Synchro{
		$t \leftarrow$ next RDF triple matching $tp$ in $\mathcal{D}$ \;
		\textbf{let} $\mu \leftarrow$ set of solutions mappings such as $\mu(t) = tp$\;
		\Return $\mu$
  	}
  }

  \Fn{Save()}{
    \Return $\langle tp, id(t) \rangle$ \;
  }

  \Fn{Load($id$)}{
	$t \leftarrow$ $\mathcal{I}_{tp}$.Locate($id$) \tcc*[f]{Locate the last triple read}\;
  }
\end{algorithm}

\paragraph{\textbf{Basic Graph patterns (AND):}}

A Basic Graph pattern (BGP) evaluation corresponds to the natural join of a set of triple patterns.
Join operators fall into three categories \cite{garcia2008database}:
\begin{inparaenum}
	\item Hash-based joins \emph{e.g.}, Symmetric Hash join or XJoin,
	\item Sort-based joins, \emph{e.g.}, (Sort) Merge join,
	\item Loop joins, \emph{e.g.}, index loop or nested loop joins.
\end{inparaenum}

Hash-based joins~\cite{graefe1993query,wilschut1993dataflow} operators
are not suitable for preemption.  As they build an in-memory hash
table on one or more inputs to evaluate the join, they are
considered as full-mappings operators. The sort-based joins and loop
joins are mapping-at-a-time operators; consequently,  they could be preempted with
low overhead.  However, for sort-based joins, we can only consider
merge joins, \emph{i.e.}, joins inputs are already sorted on the join
attribute. Otherwise, this will require to perform an in-memory sort on
the inputs. In the following, we present algorithms for building preemptable
Merge join and preemptable Index Loop join operators:

\begin{algorithm}
  \SetAlgoVlined
  \SetKwInput{Data}{Require}
  \SetKwBlock{Synchro}{non interruptible}{{end synchronized}}
  \SetKwFor{Foreach}{for each}{do}{endfor}
  \SetKwProg{Fn}{Function}{\string:}{}
  \SetKwProg{Proc}{Procedure}{\string:}{}
  \SetKwProg{Event}{Event}{\string:}{}
  \SetKwComment{tcp}{}{}%
  \SetKwComment{tcc}{// }{}%

  \DontPrintSemicolon
  \SetInd{0.55em}{0.55em}
  \caption{A \textit{Preemptable Merge Join Iterator} $I$,
  joining the output of two iterators $I_{\text{left}}$ and $I_{\text{right}}$.}
  \label{algo:merge_join}

  \Data{
    $I_{\text{left}}$: the outer join input,
    $I_{\text{right}}$: the inner join input,
    $\mu_l$: the last element read from $I_{\text{left}}$,
    $\mu_r$: the last element read from $I_{\text{right}}$,
    $\mathcal{D}$: RDF dataset queried
  }

   \vspace{-0.4cm}
  \begin{multicols}{2}

  \Fn{Stop()}{
    $I_{\text{left}}$.Stop() \;
    $I_{\text{right}}$.Stop() \;
  }

  \Fn{Save()}{
    \textbf{let} $s_l \leftarrow$ $I_{\text{left}}$.Save() \;
    \textbf{let} $s_r \leftarrow$ $I_{\text{right}}$.Save() \;
    \Return $\langle s_l, s_r, \mu_l, \mu_r \rangle$ \;
  }

  \Fn{Load($s_l, s_r, \mu'_l, \mu'_r$)}{
  $I_{\text{left}}$.Load($s_l$) \;
  $I_{\text{right}}$.Load($s_r$) \;
  $\mu_l \leftarrow \mu'_l$ \;
  $\mu_r \leftarrow \mu'_r$ \;
}

  \end{multicols}    \vspace{-0.2cm}
\end{algorithm}

\begin{algorithm}
  \SetAlgoVlined
  \SetKwInput{Data}{Require}
  \SetKwBlock{Synchro}{non interruptible}{{end synchronized}}
  \SetKwFor{Foreach}{for each}{do}{endfor}
  \SetKwProg{Fn}{Function}{\string:}{}
  \SetKwProg{Proc}{Procedure}{\string:}{}
  \SetKwProg{Event}{Event}{\string:}{}
  \SetKwComment{tcp}{}{}%
  \SetKwComment{tcc}{// }{}%

  \DontPrintSemicolon
  \SetInd{0.55em}{0.55em}
  \caption{A \textit{Preemptable Index Join Iterator} $I_i$:
  a preemptable join operator used by \sagee for BGP evaluation}
  \label{algo:bgp_iterator}

  \Data{
  	$I_{\text{left}}$: the iterator responsible for the evaluation of the outer join input,
    $tp_r$: the inner join input,
    $\mathcal{D}$: RDF dataset queried.
  }

  \Fn{Open()}{
    $I_{\text{left}}.Open()$ \;
    $\mu_c \leftarrow nil$ \;
    $I_{\text{find}} \leftarrow$ a \emph{PreemptableIndexScanIterator} over $\emptyset$\;
  }

  \Fn{GetNext()}{
    \While{$\neg I_{\text{find}}.HasNext()$}{ \label{algo:line:start_outer}
      $\mu_c \leftarrow I_{\text{left}}.GetNext()$ \; \label{algo:line:pull_mappings}
      \If{$\mu_c =  nil $}{
        \Return $nil $ \;
      }
      $I_{\text{find}} \leftarrow$ \emph{PreemptableIndexScanIterator} over
      $\llbracket \mu_c(tp_r) \rrbracket_\mathcal{D}$\; \label{algo:line:get_cursor} \label{algo:line:outer_end}
    }
    \Synchro{ \label{algo:line:start_synchro}
      \textbf{let} $\mu \leftarrow I_{\text{find}}.GetNext()$ \;
      \Return $\mu \cup \mu_c$ \; \label{algo:line:pull_boundpattern}
    }
  }

   \Fn{Load($tp', \mu', t$)}{
    $tp_r \leftarrow tp'$ \;
    \If{$\mu' \neq nil $}{
      $\mu_c \leftarrow \mu'$ \;
      $I_{\text{find}} \leftarrow$ \emph{PreemptableIndexScanIterator} over
      $\llbracket \mu_c(tp_i) \rrbracket_\mathcal{D}$\;
      $I_{\text{find}}$.Load($t$); \label{algo:line:offset}
    }
  }
\vspace{-0,5cm}
  \begin{multicols}{2}
  \Fn{Save()}{
    \textbf{let} $t \leftarrow$ the last triple read by $I_{\text{find}}$ \;
    \Return $\langle tp_r, \mu_c, t \rangle$ \;
  }
  \Fn{Stop()}{
    $I_{\text{left}}.Stop()$ \;
    $I_{\text{find}}.Stop()$ \;
  }
  \end{multicols}\vspace{-0,2cm}
\end{algorithm}

\begin{asparadesc}
\item[Preemptable merge join:]\label{sec:merge_join}

The Merge join algorithm merges the solutions mappings from two join
inputs, \emph{i.e.}, others operators sorted on the join attribute.
\sagee extends the classic merge join \cite{graefe1993query} to the
\emph{Preemptable Merge join} iterator as shown in Algorithm~\ref{algo:merge_join}.  Basically, its \texttt{Stop},
\texttt{Save} and \texttt{Load} functions recursively stop, save or
load the joins inputs, respectively. Thus, the only internal data
structures holds by the join operator are the two inputs and the last
sets of solution mappings read from them.  As described in
Table~\ref{tab:complex}, the local state of a merge join is resumable
in constant time, while its space complexity depends on the size of
two sets of solution mappings saved.

\item[Preemptable Index Loop join:]\label{sec:index_join}
The Index Loop join algorithm \cite{graefe1993query} exploits indexes
on the inner  triple pattern for efficient join processing. This algorithm has already been
used for evaluating BGPs in~\cite{olaf2009}.
\sagee extends the classic Index Loop joins to  a \emph{Preemptable
  Index join Iterator} (PIJ-Iterator) presented  in
  Algorithm~\ref{algo:bgp_iterator}.  To produce solutions, the iterator
executes the same steps repeatedly until all solutions are produced:
\begin{inparaenum}[(1)]
  \item It pulls solutions mappings $\mu_c$ from its predecessor.
  \item It applies $\mu_c$ to $tp_i$ to generate a \emph{bound pattern} $b = \mu_c(tp_i)$.
  \item If $b$ has no solution mappings in $\mathcal{D}$,
  it tries to read again from its predecessor (jump back to Step 1).
  \item Otherwise, it reads triple matching $b$ in $\mathcal{D}$, produces the associated set of solution mappings and then goes back to Step 1.
\end{inparaenum}

A \emph{PIJ-Iterator} supports preemption through the \texttt{Stop},
\texttt{Save} and \texttt{Load} functions.  The non-interruptible section of
\texttt{GetNext()} only concerns a scan in the dataset. Therefore, in the
worse case, the iterator has to wait for one index scan to complete
before being interruptible.  As with the merge join, the \texttt{Save}
and \texttt{Load} functions needs to stop and resume the left join
input. The latter also needs to resume an Index Scan, which can be
resumed in $\mathcal{O}(\log_b(|\mathcal{D}|)$.

Regarding the saved state, the iterator saves the triple pattern
joined with the last set of solution mappings pulled from the
predecessor and an index scan.  For instance, in  Figure~\ref{fig:sage_page},
the saved state of the join of $tp_2$ and $tp_1$ is $\mu_c$ {\small = \texttt{\{?v1
  $\mapsto$ wm:Product12, ?v3 $\mapsto$ "4356"\}}}.
\end{asparadesc}

\paragraph{\textbf{UNION fragment:}} The UNION fragment is defined as
the union of solution mappings from two graph patterns.
We consider a \emph{multi-set union} semantic, as set unions require
to save intermediate results to remove duplicates and thus cannot be
implemented as mapping-at-a-time operators. The set semantic can
be restored by the smart Web client using the DISTINCT modifier on
client-side.  Evaluating a multi-set union is equivalent to the
\emph{sequential evaluation} of all graph patterns in the union.
When preemption occurs, all graph patterns in the union are saved
recursively. So the union has no local state on its own.
Similarly, to resume an union evaluation, each graph
pattern is reloaded recursively.

\paragraph{\textbf{FILTER fragment:}} A SPARQL FILTER is denoted
$F = \sigma_\mathcal{R}(P)$, where $P$ is a graph pattern and
$\mathcal{R}$ is a built-in filter condition.  The evaluation of $F$
yields the solutions mappings of $P$ that verify $\mathcal{R}$.  Some
filter conditions require collection of mappings to be evaluated, like
the EXISTS filter which requires the evaluation of a graph pattern.
Consequently, we limit the filter condition to \emph{pure logical
  expressions} ($=, <, \geq, \wedge,$ etc) as defined
in~\cite{perez2009semantics,Schmidt0L10}.  The preemption of a FILTER
is similar to those of a projection. We only need to suspend or resume
$P$, respectively. For the local state, we need to save the last
solution mappings pulled from $P$ and $\mathcal{R}$.

Table~\ref{tab:complex} summarizes the complexities of \texttt{Suspend} and
\texttt{Resume} in time and space. The space complexity to \texttt{Save}  a
physical plan is established to $O(|Q| \times \log_2(|\mathcal{D}|)$. We supposed that
$|var(P)|$ and the number of $|tp|$ to save are close to $|Q|$.
However, the size of index IDs are integers that can be as big as the number of RDF triples
in $\mathcal{D}$. Hence, they can be encoded in at most $\log_2(|\mathcal{D}|)$ bits.
The time complexity to \texttt{Resume} a physical plan is
$\mathcal{O}(|Q|\times\log_b(|\mathcal{D}|))$. It is higher than the
time complexity of \texttt{Suspend}, as resuming index scans can be costly.
Overall, the complexity of Web preemption is higher than the initial $O(|Q|)$
stated in Section~\ref{sec:web-preemtion-model}. However, we demonstrate empirically in Section~\ref{sec:exp_study}
 that time and space complexity can  be kept under a few milliseconds and kilobytes, respectively.

\subsection{\sagee smart Web client}
\label{sec:sagee-smart-web}

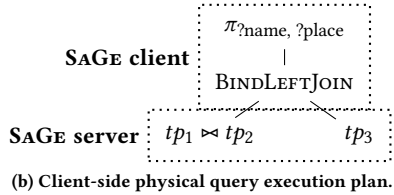
\begin{figure}
  \centering
  \subfloat [SPARQL query $Q_3$: finds all actors with their names and  their birth places, if they exist.] {\label{query:q3}
  \resizebox{0.45\textwidth}{!}{%
\begin{lstlisting}[
       basicstyle=\scriptsize\sffamily,
       language=sparql,
       extendedchars=true
    ]
PREFIX rdfs: <http://www.w3.org/2000/01/rdf-schema#>
PREFIX dbo: <http://dbpedia.org/ontology/>
SELECT ?name ?place WHERE {
 ?actor a dbo:Actor . # tp1
 ?actor rdfs:label ?name . # tp2
 OPTIONAL { ?actor dbo:birthPlace ?place . # tp3 }
}
\end{lstlisting}
  }
  }\vfill
  \subfloat [Client-side physical query execution plan.] {\label{fig:sage_client_plan}
    \resizebox{0.3\textwidth}{!}{%
     \begin{tikzpicture}
 \tikzstyle{level 1}=[level distance=0.7cm, sibling distance=2cm]
 \tikzstyle{level 2}=[sibling distance=2cm]
 \tikzstyle{level 3}=[sibling distance=2cm]
  \node (proj) {$\pi_{\text{?name, ?place}}$}
  child {
    	node (bind_join) {\textsc{BindLeftJoin}}
	child { node (bgp) {$tp_1 \bowtie tp_2$} }
	child { node (tp3) {$tp_3$} }
  };

\node[draw,thick,dotted,fit=(proj) (bind_join),label=left:{\textbf{\sagee client}}] {};
\node[draw,thick,dotted,fit=(bgp) (tp3),label=left:{\textbf{\sagee server}}] {};

\end{tikzpicture}
    }
  }
  \caption{Physical query execution plan used by the \sagee smart Web client for the  query $Q_3$.}
  \label{fig:client_plan}
\end{figure}

The \sagee approach requires a \emph{smart Web client} for processing SPARQL
queries for two reasons.
First, the smart Web client must be able to
  continue query execution after receiving a saved plan from the server.
Second, as the preemptable server only implements a fragment of
  SPARQL, the smart Web client has to implement the missing operators
  to support full SPARQL queries.  It includes SERVICE, ORDER BY,
  GROUP BY, DISTINCT, MINUS, FILTER EXIST and aggregations
  (COUNT, AVG, SUM, MIN, MAX), but also
  $\bowtie, \cup, \leftouterjoin$ and $\pi$ to be able to recombine
  results obtained from the \sagee server.
Consequently, the \sagee smart client is a \emph{SPARQL query engine}
that accesses RDF datasets through the \sagee server. Given a SPARQL
query, the \sagee client parses it into a logical query execution
plan, optimizes it and then builds its own physical query
execution. The leafs of the plan must correspond to subqueries
evaluable by a \sagee server. Figure \ref{fig:sage_client_plan} shows the execution plan
build by the \sagee client for executing query $Q_3$, from Figure \ref{query:q3}.

Compared to a SPARQL endpoint, processing SPARQL queries with the
\sagee smart client has an overhead in terms of the number of request
sent to the server and transferred data\footnote{The client overhead
  should not be confused with the server overhead.}. With a SPARQL endpoint, a web
client executes a query by sending a single web request and receives
only query results. The smart client overhead for executing the same
query is the additional number of request and data transfered to
obtain the same results. Consequently, the client overhead has two
components:
\begin{asparadesc}
\item[Number of requests:] To execute a SPARQL query, a \sagee client
  needs $n \geq 1$ requests.  We can distinguish two cases:
  \begin{inparaenum}
  \item The query is fully supported by the \sagee server, so
    $n$ is equal to the number of time quantum required to process the
    query. Notice that the client needs to pay the network latency
    twice per request.
  \item The query is not fully supported by the \sagee server. Then,
    the client decomposes the query into a set of subqueries supported by
    the server, evaluate each subquery as in the first case, and
    recombine intermediate results to produce query results.
  \end{inparaenum}

\item[Data transfer:] We also distinguish two cases:
  \begin{inparaenum}
  \item If the query is fully supported by the \sagee server, the only
    overhead is the size of the saved plan $S_i$ multiplied by the
    number of requests. Notice that the saved plan $S_i$ can be returned by
    reference or by value, \emph{i.e.}, saved server-side or client-side.
  \item If the query is not fully supported by the \sagee server, then
    the client decomposes the query and recombine results of
    subqueries. Among these results, some are
    intermediate results and part of the client overhead.
   \end{inparaenum}
\end{asparadesc}

Consequently, the challenge for the smart client is to \emph{minimize
the overhead in terms of the number of requests and transferred data.}
The data transfer could be reduced by saving the plans server-side rather
than returning it to clients\footnote{In our implementation, we choose to save plans client-side to tolerate server failures.}.
However, the main source of client overhead comes from the decomposition
made to support full SPARQL queries, as these queries increase the number of requests sent to the server.
Some decompositions are more costly than others.
To illustrate, consider the evaluation of
$(P_1 \text{ OPTIONAL } P_2)$ where $P_1$ and $P_2$ are expressions
supported by the \sagee server.  A possible approach is to evaluate
$\llbracket P_1 \rrbracket_\mathcal{D}$ and
$\llbracket P_2 \rrbracket_\mathcal{D}$ on the server, and then
perform the left outer join on the client. This strategy generates
only two subqueries but materializes
$\llbracket P_2 \rrbracket_\mathcal{D}$ on client. If there is no join
results, $\llbracket P_2 \rrbracket_\mathcal{D}$ are just useless
intermediate results.

Another approach is to rely on a \emph{nested loop join approach}:
evaluates $\llbracket P_1\rrbracket_\mathcal{D}$ and for each
$\mu_1 \in \llbracket P_1\rrbracket_\mathcal{D}$, if
$\mu_2 \in \llbracket \mu_1(P2) \rrbracket_\mathcal{D}$ then
$\{ \mu_1 \cup \mu_2 \}$ are solutions to $P_1\leftouterjoin P_2$.
Otherwise, only $\mu_1$ is a solution to the left-join. This approach
sends \emph{at least} as many subqueries to the server than there is solutions to
$\llbracket P_1\rrbracket_\mathcal{D}$.

To reduce the communication, the \sagee client implements
\textsc{Bind-LeftJoins} to process local join in a block fashion, by sending
unions of BGPs to the server.  This technique is already used in
federated SPARQL query processing~\cite{schwarte2011fedx} with bound
joins and in BrTPF~\cite{hartig2016bindings}. Consequently, the number
of request to the \sagee server is reduced by a factor equivalent to
the size of a block of mappings.  However, the number of requests sent still depends
on the cardinality of $P_1$.

Consequently, we propose a new technique, called \textsc{OptJoin}, for optimizing
the evaluation of a subclass of left-joins. The approach relies on the
fact that:
\[
  \llbracket P_1 \leftouterjoin P_2 \rrbracket_\mathcal{D} =
  \llbracket P_1 \bowtie P_2 \rrbracket_\mathcal{D}
  \cup
  (  \llbracket  P_1 \rrbracket_\mathcal{D} \setminus  \llbracket \pi_{var(P_1)}(P_1 \bowtie P_2)
  \rrbracket_\mathcal{D})
\]
So we can deduce that:
$
  \llbracket P_1 \leftouterjoin P_2 \rrbracket_\mathcal{D}
  \subseteq  \llbracket (P_1 \bowtie P_2 )  \cup P_1 \rrbracket_\mathcal{D}
  $.
If both $P_1$ and $P_2$ are evaluable by the \sagee server, then
the smart client computes the left-join as follows.  First, it
sends the query $(P_1 \bowtie P_2) \cup P_1$ to the server.  Then, for
each mapping $\mu$ received, it builds local materialized views for
$\llbracket P_1 \bowtie P_2 \rrbracket_\mathcal{D}$ and
$\llbracket P_1 \rrbracket_\mathcal{D}$.  The client knows that
$\mu \in \llbracket P_1 \bowtie P_2 \rrbracket_\mathcal{D}$ if
$\emph{dom }(\mu) \subset \emph{var }(P_1 \bowtie P_2)$, otherwise
$\mu \in \llbracket P_1 \rrbracket_\mathcal{D}$.  Finally, the client
uses the views to process the left-join locally.
With this technique, the client only use one subquery to evaluate the left-join
and, in the worst case, it transfers $\llbracket  P_1 \bowtie P_2\rrbracket_\mathcal{D}$
as additional intermediate results.

To illustrate, consider query $Q_3$ from Figure~\ref{query:q3}.
$Q_3$, with $88 334$ solutions. The cardinality of
$tp_1 \bowtie tp_2$ is also of $88 334$, as every actor has a birth
place. This is the worse case for a \textsc{BindLeftJoin}, which will
require $\tfrac{88 334}{\emph{Block size}}$ additional requests to evaluate the left join.  However,
with an \textsc{OptJoin}, the client is able to evaluate $Q_3$ in approximately $500$ requests.

We implement both \textsc{BindLeftJoin} and \textsc{OptJoin} as physical
operators to evaluate OPTIONALs, depending on the query.
We also implement regular \textsc{BindJoin} for processing SERVICE queries.

% -------------------------------------------------------
\section{Experimental study}\label{sec:exp_study}

We want to empirically answer the following questions: What is the
overhead of Web preemption in time and space? Does Web preemption
improves the average workload completion time? Does Web preemption
improves the time for first results? What are the client overheads
in terms of numbers of requests and data transfer? We use Virtuoso as the
baseline for comparing with SPARQL endpoints, with TPF and BrTPF as the
baselines for LDF approaches.

We implemented the \sagee client in Java, using Apache
Jena\footnote{\url{https://jena.apache.org/}}. As an extension of
Jena, \sagee is just as compliant with SPARQL 1.1.  The \sagee server
is implemented as a Python Web service and uses HDT \cite{fernandez2013binary}
(v1.3.2) for storing data. Notice that the current implementation of
HDT cannot ensure $log_b(n)$ access time for all triple patterns,
like $(?s \, p \, ?o)$. This impacts negatively the
performance of \sagee when resuming some queries.
The code and the experimental setup are available on the
companion website\footnote{\url{https://github.com/sage-org/sage-experiments}}.

\subsection{Experimental setup}

\begin{asparadesc}
\item[Dataset and Queries:] We use the Waterloo SPARQL Diversity
  Benchmark (WatDiv\footnote{\url{http://dsg.uwaterloo.ca/watdiv/}})
  \cite{alucc2014diversified}. We re-use the RDF dataset and the SPARQL queries from the BrTPF \cite{hartig2016bindings} experimental study\footnote{\url{http://olafhartig.de/brTPF-ODBASE2016}}.  The dataset contains $10^7$
  triples and queries are arranged in 50 workloads of 193 queries each. They are SPARQL conjunctive queries with
  STAR, PATH and SNOWFLAKE shapes. They vary in complexity, up
  to $10$ joins per query with very high and very low selectivity.
  $20\%$ of queries require more than $\approx 30s$ to be executed using the
  virtuoso server. All workloads follow nearly the same distribution of query
  execution times as presented in Figure~\ref{fig:workload}. The
  execution times are measured for one workload of 193 queries with
  \sagee and an infinite time quantum.

 \begin{figure}
 	\centering
	\includegraphics[width=0.3\textwidth]{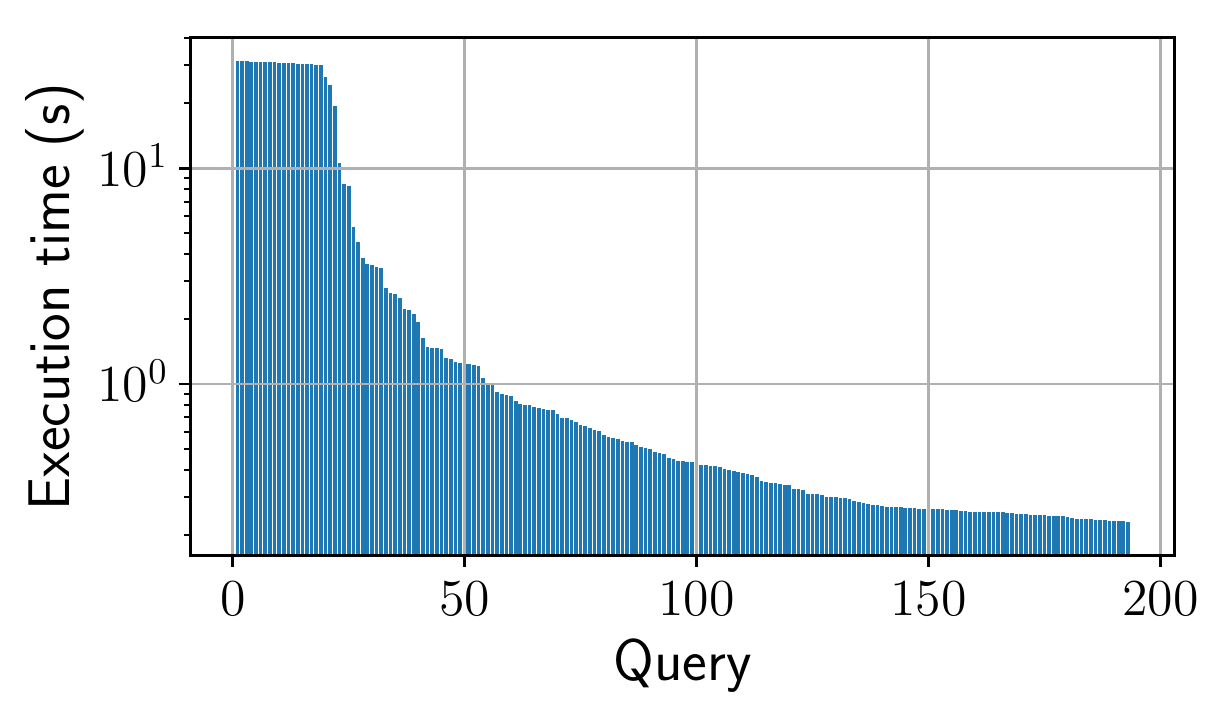}
	\caption{Distribution of query execution time.}
	\label{fig:workload}
 \end{figure}

\item[Approaches:] We compare the following approaches:
  \begin{asparaitem}
  \item \sage: We run the \sagee query engine with various time quantum: 75ms and 1s, denoted \sage-75ms and \sage-1s respectively. HDT indexes are loaded in memory while HDT data is stored on disk.

  \item \emph{Virtuoso:} We run the Virtuoso SPARQL
    endpoint~\cite{ErlingM09} (v7.2.4), \emph{without any quotas or limitations} .

  \item \emph{TPF:} We run the standard TPF client (v2.0.5) and TPF
    server (v2.2.3) with HDT files as backend (same settings as \sagee).

  \item \emph{BrTPF:} We run the BrTPF client and server used in \cite{hartig2016bindings}, with HDT files as backend (same settings as \sagee).  BrTPF
    is currently the LDF approach with the lowest data transfer
    \cite{hartig2016bindings}.
  \end{asparaitem}

\item[Servers configurations:] We run all the servers on a machine
  with Intel(R) Xeon(R) CPU E7-8870@2.10GHz and 1.5TB RAM.

\item[Clients configurations:] In order to generate load over servers,
  we rely on 50 clients, each one executing a different workload of
  queries.  All clients start executing their workload simultaneously.
  The clients access servers through HTTP proxies to ensure that
  client-server latency is kept around 50ms.

\item[Evaluation Metrics:] Presented results correspond to the average
  obtained of three successive execution of the queries workloads.
  \begin{inparaenum}
  \item \textit{Workload completion time (WCT)}: is the total time
    taken by a client to evaluate a set of SPARQL queries, measured as
    the time between the first query starting and the last query
    completing.
  \item \textit{Time for first results (TFR)} for a query: is  the
    time between the query starting and the production of the first
    query's results.
  \item \textit{Time preemption overhead}: is the total time taken by the
    server's \texttt{Suspend} and \texttt{Resume} operations.
  \item \textit{Number of HTTP requests and data transfer}: is the total number of HTTP
    requests sent by a client to a server and the number of transferred data
    when executing a SPARQL query.
  \end{inparaenum}
\end{asparadesc}

\subsection{Experimental results}

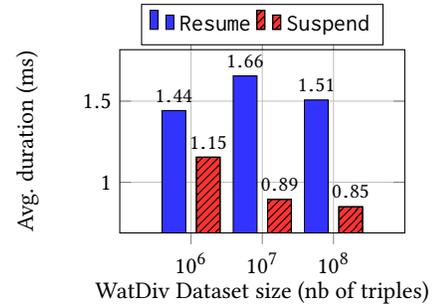
\begin{figure}
	\centering
   \begin{tikzpicture}
	\begin{axis}[
		scale=0.45,
		height=7cm,
    width=10cm,
		xlabel=WatDiv Dataset size (nb of triples),
		ylabel=Avg. duration (ms),
		enlargelimits=0.2,
		enlarge x limits=0.5,
		legend style={at={(0.5,1.25)},anchor=north},
		legend columns=2,
		ybar=4pt,
		bar width=9pt,
		symbolic x coords={$10^6$, $10^7$, $10^8$},
		xtick=data,
		every node near coord/.append style={font=\small},
		nodes near coords,
		grid=both,
    		grid style={line width=.1pt, draw=gray!10},
    		major grid style={line width=.2pt,draw=gray!50}]
		\addplot[black,fill=blue!80!white] coordinates {($10^6$, 1.442045133820285) ($10^7$, 1.6558038462400018) ($10^8$, 1.5072508498660284)};
		\addplot[black,fill=red!80!white,postaction={pattern=north east lines}] coordinates {($10^6$, 1.1547279856905164) ($10^7$, 0.894600288705793) ($10^8$, 0.8491159259080842)};
		\legend{\texttt{Resume}, \texttt{Suspend}}
	\end{axis}
\end{tikzpicture}
	\caption{Average preemption overhead.}
	\label{fig:overhead}
\end{figure}

\begin{figure}
	\centering
	\includegraphics[width=0.4\textwidth]{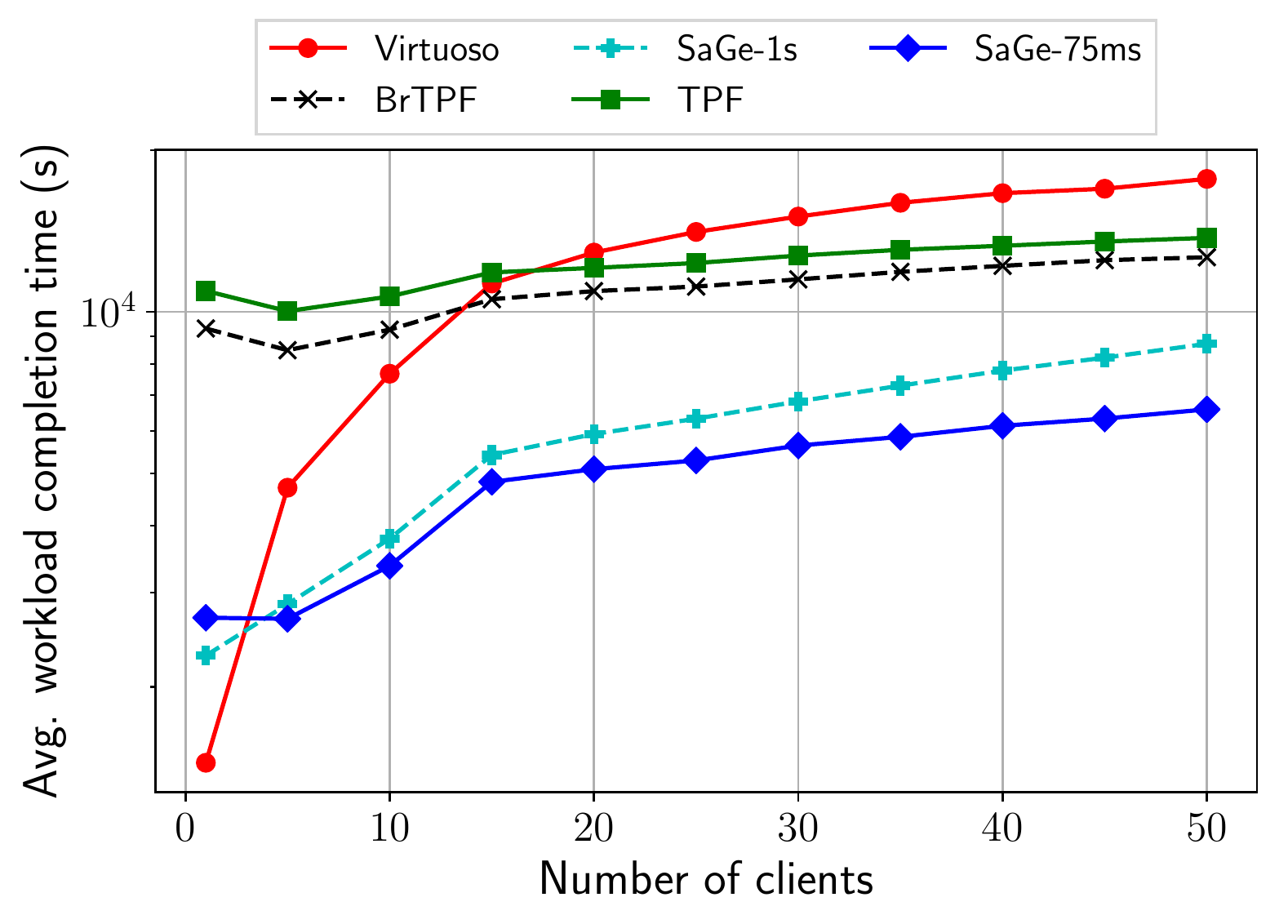}
	\caption{Average workload completion time per client, with up to 50 concurrent clients (logarithmic scale).}
	\label{fig:total_time}
\end{figure}

\begin{figure}
	\centering
	\includegraphics[width=0.4\textwidth]{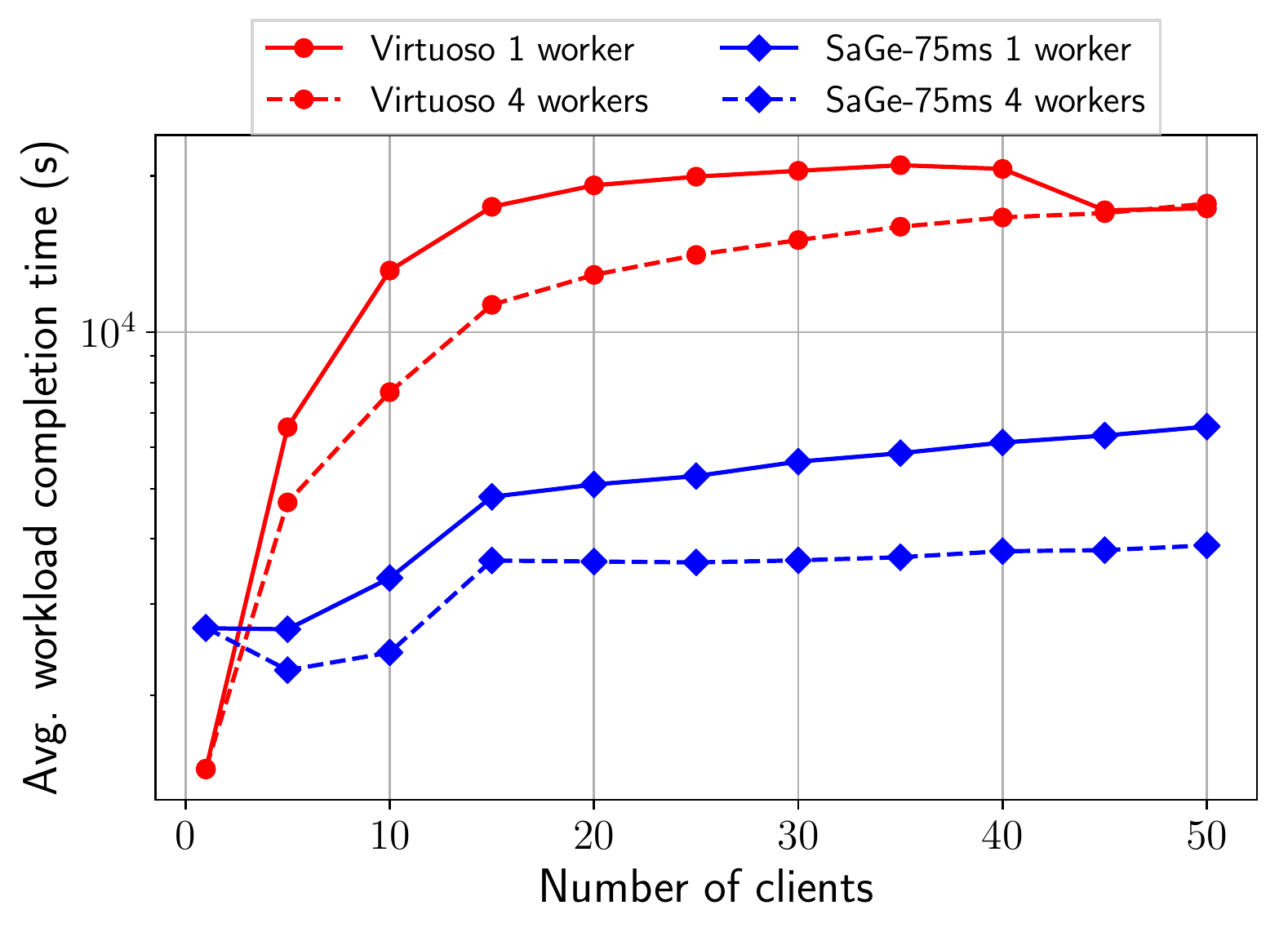}
	\caption{Average workload completion time per client, with 4 workers (logarithmic scale).}
	\label{fig:total_time_4w}
\end{figure}

\begin{figure}
	\centering
	\includegraphics[width=0.4\textwidth]{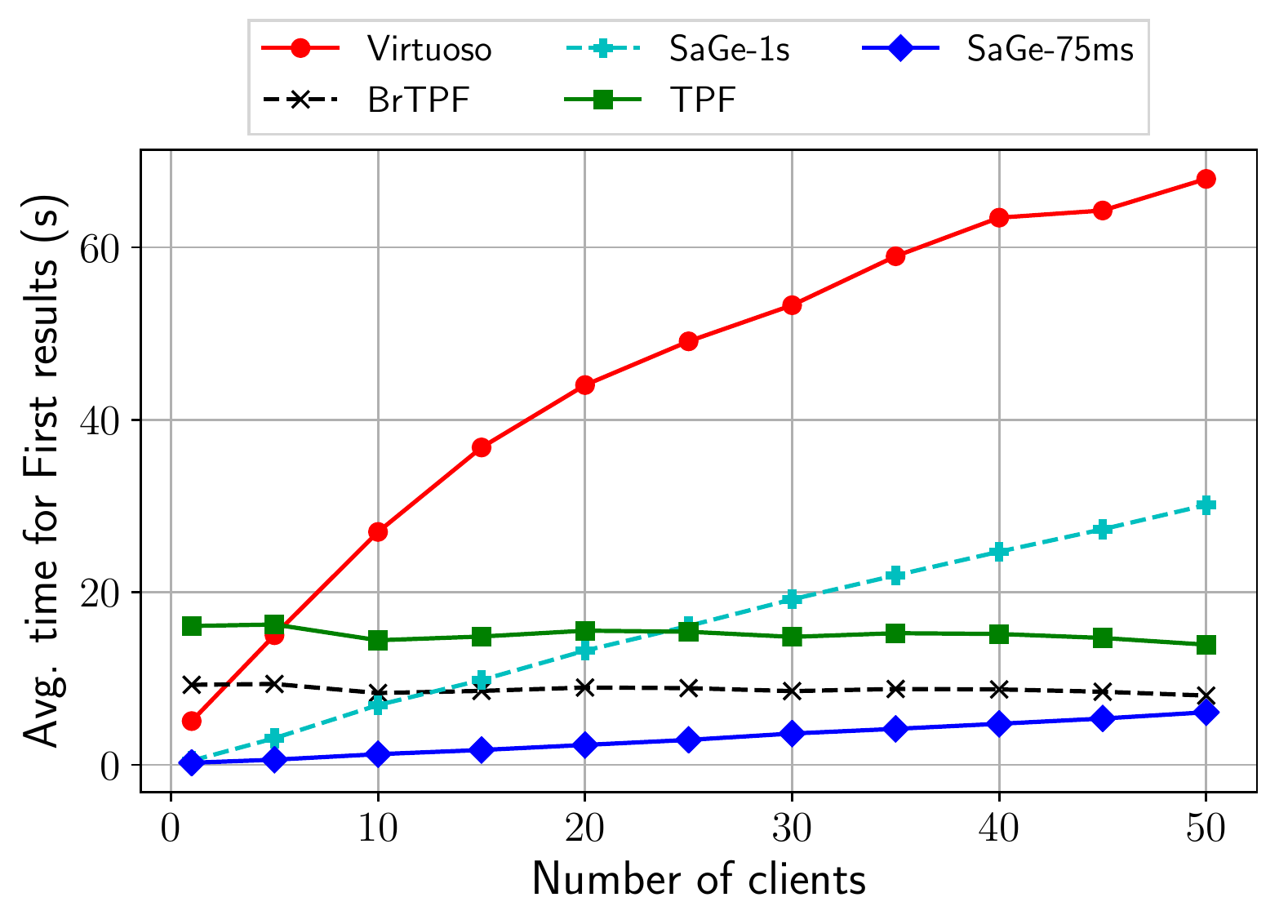}
	\caption{Average time for first results (over all queries), with up to 50 concurrent clients (linear scale).}
	\label{fig:first_results}
\end{figure}

\begin{table}
	\centering
	\begin{tabular}{|c|c|c|c|}
		\hline
		 \textbf{Mean} & \textbf{Min} & \textbf{Max} & \textbf{Standard deviation} \\
		 \hline
		  $1.716$ kb & $0.276$ kb & $6.212$ kb & $1.337$ kb \\
		  \hline
	\end{tabular}
	\caption{Space of saved physical query execution plans.}
	\label{table:space_overhead}
\end{table}

We first ensure that the \sagee approach yield complete results. We
run both Virtuoso  and \sagee and verify
that, for each query, \sagee delivers complete results using Virtuoso
results as ground truth.

\paragraph{What is the overhead in time of Web preemption?}

The overhead in time of Web preemption is the time spent by
the Web server for suspending a running query and the time spent
for resuming the next waiting query.
To measure the overhead, we run one workload of queries using \sage-75ms and
measure time elapsed for the \texttt{Suspend} and \texttt{Resume} operations.
Figure~\ref{fig:overhead} shows the overhead in time for \sagee
\texttt{Suspend} and \texttt{Resume} operations using different sizes of
the WatDiv dataset.  Generally, the size of the dataset does not
impact the overhead, which is around $1ms$ for \texttt{Suspend} and
$1.5ms$ for \texttt{Resume}. As expected, the overhead of the
\texttt{Resume} operation is greater than the one of the
\texttt{Suspend} operation, due to the cost of resuming the Index
scans in the plan. With a quantum of 75ms, the overhead is $\approx3\%$ of the
quantum, which is negligible.

\paragraph{What is the overhead in space of Web preemption?}

The overhead in space of the Web preemption is the size of
saved plans produced by the \texttt{Suspend} operation. According to
Section~\ref{sec:sagee-preempt-serv}, we determined that the \sagee
physical query plans can be saved in $O(|Q|)$.
To measure the overhead, we run a workload of queries using \sage-75ms and
measure the size of saved plans. Saved plans are compressed
using Google Protocol Buffers~\footnote{\url{https://developers.google.com/protocol-buffers/}}.
Table~\ref{table:space_overhead} shows the overhead in space for
\sage.  As we can see the size of a saved plan remains very small,
with no more than 6 kb for a query with ten joins.  Hence, this space
is indeed proportional to the size of the plan suspended.

\paragraph{Does Web preemption improve the average workload
  completion time?}

To enable Web preemption, the \sagee server has a restricted choice of
physical query operators, thus physical query execution plans
generated by the \sagee server should be less performant than those
generated by Virtuoso.  This tradeoff only make sense if the Web
preemption compensates the loss in performance. Compensation is
possible only if the workload alternates long-running and short
running queries.
In the setup, each client runs a different workload of 193 queries that
vary from $30$s to $0.22$s, following an exponential distribution.  All clients
execute their workload concurrently and start simultaneously. We
experiment up to 50 concurrent clients by step of 5 clients.
As there is only one worker on the Web server, and queries
execution time vary, this setup is the worst case for Virtuoso.

Figure~\ref{fig:total_time} shows the average workload completion time
obtained for all approaches, \emph{with a logarithmic scale}.
As expected, Virtuoso is significantly impacted by the convoy effect and
delivers the worse WTC after 20 concurrent clients. TPF and BrTPF
avoid the convoy and behave similarly. BrTPF is more performant
thanks to its bind-join technique that group requests to the server.
\sage-75ms and \sage-1s avoid the convoy effect and delivers better WTC
than TPF and BrTPF.  As expected, increasing the time quantum also increases the
probability of convoy effect and, globally, \sage-75ms offers the best WTC.
We rerun the same experiment with 4 workers for \sage-75ms and Virtuoso.
Figure~\ref{fig:total_time_4w} shows the average workload completion
time obtained for both approaches. As we can see, both approaches
benefit of the four workers. However, Virtuoso still suffers from the
convoy effect.

\paragraph{Does Web preemption improve the time for the first results?}

The Time for first results (TFR) for a query is the time
between the query starting and the production of the first query's
results.
Web preemption should provides better time for first results. Avoiding
convoy effect allows to start queries earlier and then get results
earlier. We rerun the same setup as in the  previous section and measure the time
for first results (TFR). Figure~\ref{fig:first_results} shows the results \emph{with a linear
scale}.
As expected, Virtuoso suffers from the convoy effect that degrades
significantly the TFR when the concurrency increases.
TPF and BrTPF do not suffer from the convoy effect and TFR is clearly
stable over concurrency. The main reason is that delivering a page of
result for a single triple pattern takes $\approx5$ms in our experiment, so
the waiting time on the TPF server grows very slowly. BrTPF is better
than TPF due to its bind-join technique.
The TFR for \sage-75ms and \sage-1s increases with the number of clients
and the slope seems proportional to the quantum. This is normal
because the waiting time on the server increases with the number of clients, as seen previously.
Reducing the quantum improves the TFR, but increases the
number of requests and thus deteriorates the WTC.

\paragraph{What are the client overheads in terms of number of requests
  and data transfer?}

\begin{table}
	\subfloat[Results with WatDiv and FEASIBLE-DBpedia datasets\label{table:http_requests}]{
		\resizebox{0.48\textwidth}{!}{%
	\begin{tabular}{|c|c|c|c|c|c|}
		\hline
		 \textbf{Dataset} & \textbf{Virtuoso} & \textbf{\sage-1s} & \textbf{\sage-75ms} & \textbf{BrTPF} & \textbf{TPF} \\
		 \hline
		 WatDiv $10^7$ & 193 & 645 & 4 082 & $9,2 \cdot 10^4$ & $2,55 \cdot 10^5$ \\
		 \hline
		 FEASIBLE & 166 & 1 822 & 3 305 & $2,95 \cdot 10^4$ & $1,86 \cdot 10^5$ \\
		  \hline
	\end{tabular}
	}
	}\hfill
	\subfloat[Comparison for \textsc{BindLeftJoin} and \textsc{OptJoin} operators\label{table:optional}]{
	\begin{tabular}{|c|c|c|}
		\hline
		 \textbf{Time quantum} & \textbf{\sage+\textsc{BindLeftJoin}} & \textbf{\sage+\textsc{OptJoin}} \\
		 \hline
		  75ms & 72 489 & 5 656 \\
		  \hline
		  1s & 70 964 & 511 \\
		  \hline
	\end{tabular}
	}
	\caption{Average number of HTTP requests sent to server}
  \label{table:requests_optional}
\end{table}

The client overhead in requests is the number of requests  that the smart
client sent to the  server to get complete results minus one, as
Virtuoso executes all queries in one request.
As WatDiv queries are pure conjunctive queries and supported by
the \sagee server, the number of requests to the server is the number of
time quantum required to evaluate the whole workload.
We measure the number of requests sent to servers with one workload for
all approaches, with results shown in Table~\ref{table:requests_optional}\subref{table:http_requests}.
As expected, Virtuoso just send 193 requests
to the server. \sage-75ms send 4082 requests to the server, while
TPF send $2.55 \times 10^5$ requests to the TPF server. We can see also that
increasing the time quantum significantly reduces the number of requests;
\sage-1s send only 645 requests. However, this seriously deteriorates the
average WCT and TFR as presented before.
To compute the overhead in data transfer of \sage, we just need to
multiply the number of requests by the average size of saved plans;
for \sage-75ms, the client overhead in data transfer is $4082\times1,3$
kb $= 5.45$ Mo. As the total size of results is 51Mo, the client
overhead in data transfer is $\approx10\%$.  For TPF, the average size of a
page is 7ko; $2.5 \cdot 10^5 \times7k = 1.78$ Go, so the data transfer overhead
is $\approx340\%$ for TPF.

\paragraph{What are the client overheads in terms of numbers of requests
  and data transfer for more complex queries?}

The WatDiv benchmark does not generate queries with OPTIONAL or
FILTER operators. If some filters are supported by the \sagee server,
the OPTIONAL operator and other filters clearly impact the number of
requests sent to the \sagee server, as explained in
Section~\ref{sec:sagee-smart-web}.
First, we run an experiment to measure the number of requests for
queries with OPTIONAL. We generate new WatDiv queries from one set of 193
queries, using the following protocol. For each query
$Q = \{ tp_1, \dots, tp_n \}$, we select $tp_k \in Q$ with the highest
cardinality, then we generate
$Q' = tp_k \leftouterjoin (Q \setminus tp_k)$. Such queries verify
conditions for using \textsc{BindLeftJoin} and  \textsc{OptJoin}. They are
challenging for the \textsc{BindLeftJoin} as they generate many
mappings; they are also challenging for the \textsc{OptJoin} as all joins
yield results.  Table~\ref{table:requests_optional}\subref{table:optional} shows the
results when evaluating OPTIONAL queries with \textsc{OptJoin} and
\textsc{BindLeftJoin} approaches.  We observe that, in general,
\textsc{OptJoin} outperforms \textsc{BindLeftJoin}. Furthermore,
\textsc{OptJoin} is improved when using a higher quantum, as the
single subquery sent is evaluated more quickly. This is
not the case for \textsc{BindLeftJoin}, as the number of requests
still depends on the number of intermediate results.

Finally, we re-use 166 SELECT queries from \textsc{Feasible} \cite {SaleemMN15} with
the DBpedia 3.5.1 dataset, which were generated from real-users queries.
We excluded queries that time-out, identified in \cite
{SaleemMN15}, and measure the number of requests sent to the server.
Table~\ref{table:requests_optional}\subref{table:http_requests} shows the results.
Of course, Virtuoso just send 166 requests to the
server. As we can see the ratio of requests between \sage-75 and TPF
is nearly the same as the previous experiment. However, the
difference between \sage-1s and \sage-75ms has clearly decreased, because most
requests sent are produced by the decomposition of OPTIONAL and FILTER
and not by the evaluation of BGPs.

% -------------------------------------------------------
\section{Conclusion and Future Works}\label{sec:conclusion}

In this paper, we demonstrated how Web preemption allows not only to
suspend SPARQL queries, but also to resume them. This opens the
possibility to efficiently execute long-running queries with complete results.
The scientific challenge was to keep the Web preemption overhead as low as possible;
we demonstrated that the overhead of large a fragment of SPARQL can be
kept under 2ms.

Compared to SPARQL endpoint approaches without quotas, \sagee avoids the
convoy effect and is a winning bet as soon as  the queries of the workload vary in
execution time.  Compared to LDF approaches, \sagee offers a more
expressive server interface with joins, union and filter evaluated
server side. Consequently, this considerably reduces data transfer and
the communication cost, improving the execution time of SPARQL queries.

\sagee opens several perspectives.
First, for the sake of simplicity, we made the hypothesis of read-only
datasets in this paper. If we allow concurrent updates, then a query
can be suspended with a version of a RDF dataset and resumed with another
version.  The main challenge is then to determine which consistency
criteria can be ensured by the  preemptive Web server. As \sagee
only accesses the RDF dataset when scanning triple patterns, it could be
possible to compensate concurrent updates and deliver, at least,
eventual consistency for query results.
Second, we aim to explore how the interface of the server can be
extended to support named graphs and how the optimizer of the smart client can be tuned to produce
better query decompositions, especially in the case of federated SPARQL
query processing.
Third, we plan to explore if more elaborated scheduling policies could
increase performance.
Finally, the Web preemption model is not restricted to
SPARQL. An interesting perspective is to design a similar approach for
SQL or GraphQL.

\balance

\begin{acks}
This work has been partially funded through the FaBuLA project, part of the AtlanSTIC 2020 program.
\end{acks}

\bibliographystyle{ACM-Reference-Format}
\bibliography{paper}


\begin{thebibliography}{29}

%%% the Web address.

\ifx \showCODEN    \undefined \def \showCODEN     #1{\unskip}     \fi
\ifx \showDOI      \undefined \def \showDOI       #1{#1}\fi
\ifx \showISBNx    \undefined \def \showISBNx     #1{\unskip}     \fi
\ifx \showISBNxiii \undefined \def \showISBNxiii  #1{\unskip}     \fi
\ifx \showISSN     \undefined \def \showISSN      #1{\unskip}     \fi
\ifx \showLCCN     \undefined \def \showLCCN      #1{\unskip}     \fi
\ifx \shownote     \undefined \def \shownote      #1{#1}          \fi
\ifx \showarticletitle \undefined \def \showarticletitle #1{#1}   \fi
\ifx \showURL      \undefined \def \showURL       {\relax}        \fi
% invisible to TeX
\providecommand\bibfield[2]{#2}
\providecommand\bibinfo[2]{#2}
\providecommand\natexlab[1]{#1}
\providecommand\showeprint[2][]{arXiv:#2}

\bibitem[\protect\citeauthoryear{Alu{\c{c}}, Hartig, {\"{O}}zsu, and
  Daudjee}{Alu{\c{c}} et~al\mbox{.}}{2014}]%
        {alucc2014diversified}
\bibfield{author}{\bibinfo{person}{G{\"{u}}nes Alu{\c{c}}},
  \bibinfo{person}{Olaf Hartig}, \bibinfo{person}{M.~Tamer {\"{O}}zsu}, {and}
  \bibinfo{person}{Khuzaima Daudjee}.} \bibinfo{year}{2014}\natexlab{}.
\newblock \showarticletitle{Diversified Stress Testing of {RDF} Data Management
  Systems}. In \bibinfo{booktitle}{\emph{The Semantic Web - {ISWC} 2014 - 13th
  International Semantic Web Conference, Riva del Garda, Italy, October 19-23,
  2014. Proceedings, Part {I}}} \emph{(\bibinfo{series}{Lecture Notes in
  Computer Science})}, Vol.~\bibinfo{volume}{8796}.
  \bibinfo{publisher}{Springer}, \bibinfo{pages}{197--212}.
\newblock
\urldef\tempurl%
\url{https://doi.org/10.1007/978-3-319-11964-9\_13}
\showDOI{\tempurl}


\bibitem[\protect\citeauthoryear{Anderson and Dahlin}{Anderson and
  Dahlin}{2014}]%
        {RR2014}
\bibfield{author}{\bibinfo{person}{Thomas Anderson} {and}
  \bibinfo{person}{Michael Dahlin}.} \bibinfo{year}{2014}\natexlab{}.
\newblock \bibinfo{booktitle}{\emph{Operating Systems: Principles and Practice}
  (\bibinfo{edition}{2nd} ed.)}.
\newblock \bibinfo{publisher}{Recursive books}.
\newblock
\showISBNx{0985673524, 9780985673529}


\bibitem[\protect\citeauthoryear{Aranda, Hogan, Umbrich, and
  Vandenbussche}{Aranda et~al\mbox{.}}{2013}]%
        {buil2013sparql}
\bibfield{author}{\bibinfo{person}{Carlos~Buil Aranda}, \bibinfo{person}{Aidan
  Hogan}, \bibinfo{person}{J{\"{u}}rgen Umbrich}, {and}
  \bibinfo{person}{Pierre{-}Yves Vandenbussche}.}
  \bibinfo{year}{2013}\natexlab{}.
\newblock \showarticletitle{{SPARQL} Web-Querying Infrastructure: Ready for
  Action?}. In \bibinfo{booktitle}{\emph{The Semantic Web - {ISWC} 2013 - 12th
  International Semantic Web Conference, Sydney, NSW, Australia, October 21-25,
  2013, Proceedings, Part {II}}} \emph{(\bibinfo{series}{Lecture Notes in
  Computer Science})}, Vol.~\bibinfo{volume}{8219}.
  \bibinfo{publisher}{Springer}, \bibinfo{pages}{277--293}.
\newblock
\urldef\tempurl%
\url{https://doi.org/10.1007/978-3-642-41338-4\_18}
\showDOI{\tempurl}


\bibitem[\protect\citeauthoryear{Aranda, Polleres, and Umbrich}{Aranda
  et~al\mbox{.}}{2014}]%
        {axel2014}
\bibfield{author}{\bibinfo{person}{Carlos~Buil Aranda}, \bibinfo{person}{Axel
  Polleres}, {and} \bibinfo{person}{J{\"{u}}rgen Umbrich}.}
  \bibinfo{year}{2014}\natexlab{}.
\newblock \showarticletitle{Strategies for Executing Federated Queries in
  {SPARQL1.1}}. In \bibinfo{booktitle}{\emph{The Semantic Web - {ISWC} 2014 -
  13th International Semantic Web Conference, Riva del Garda, Italy, October
  19-23, 2014. Proceedings, Part {II}}} \emph{(\bibinfo{series}{Lecture Notes
  in Computer Science})}, Vol.~\bibinfo{volume}{8797}.
  \bibinfo{publisher}{Springer}, \bibinfo{pages}{390--405}.
\newblock
\urldef\tempurl%
\url{https://doi.org/10.1007/978-3-319-11915-1\_25}
\showDOI{\tempurl}


\bibitem[\protect\citeauthoryear{Bizer, Heath, and Berners{-}Lee}{Bizer
  et~al\mbox{.}}{2009}]%
        {bizer2009linked}
\bibfield{author}{\bibinfo{person}{Christian Bizer}, \bibinfo{person}{Tom
  Heath}, {and} \bibinfo{person}{Tim Berners{-}Lee}.}
  \bibinfo{year}{2009}\natexlab{}.
\newblock \showarticletitle{Linked Data - The Story So Far}.
\newblock \bibinfo{journal}{\emph{Int. J. Semantic Web Inf. Syst.}}
  \bibinfo{volume}{5}, \bibinfo{number}{3} (\bibinfo{year}{2009}),
  \bibinfo{pages}{1--22}.
\newblock
\urldef\tempurl%
\url{https://doi.org/10.4018/jswis.2009081901}
\showDOI{\tempurl}


\bibitem[\protect\citeauthoryear{Blasgen, Gray, Mitoma, and Price}{Blasgen
  et~al\mbox{.}}{1979}]%
        {BlasgenGMP79}
\bibfield{author}{\bibinfo{person}{Mike~W. Blasgen}, \bibinfo{person}{Jim
  Gray}, \bibinfo{person}{Michael~F. Mitoma}, {and} \bibinfo{person}{Thomas~G.
  Price}.} \bibinfo{year}{1979}\natexlab{}.
\newblock \showarticletitle{The Convoy Phenomenon}.
\newblock \bibinfo{journal}{\emph{Operating Systems Review}}
  \bibinfo{volume}{13}, \bibinfo{number}{2} (\bibinfo{year}{1979}),
  \bibinfo{pages}{20--25}.
\newblock
\urldef\tempurl%
\url{https://doi.org/10.1145/850657.850659}
\showDOI{\tempurl}


\bibitem[\protect\citeauthoryear{Comer}{Comer}{1979}]%
        {Comer79}
\bibfield{author}{\bibinfo{person}{Douglas Comer}.}
  \bibinfo{year}{1979}\natexlab{}.
\newblock \showarticletitle{The Ubiquitous B-Tree}.
\newblock \bibinfo{journal}{\emph{{ACM} Comput. Surv.}} \bibinfo{volume}{11},
  \bibinfo{number}{2} (\bibinfo{year}{1979}), \bibinfo{pages}{121--137}.
\newblock
\urldef\tempurl%
\url{https://doi.org/10.1145/356770.356776}
\showDOI{\tempurl}


\bibitem[\protect\citeauthoryear{Erling and Mikhailov}{Erling and
  Mikhailov}{2009}]%
        {ErlingM09}
\bibfield{author}{\bibinfo{person}{Orri Erling} {and} \bibinfo{person}{Ivan
  Mikhailov}.} \bibinfo{year}{2009}\natexlab{}.
\newblock \showarticletitle{{RDF} Support in the Virtuoso {DBMS}}.
\newblock In \bibinfo{booktitle}{\emph{Networked Knowledge - Networked Media -
  Integrating Knowledge Management, New Media Technologies and Semantic
  Systems}}. \bibinfo{pages}{7--24}.
\newblock
\urldef\tempurl%
\url{https://doi.org/10.1007/978-3-642-02184-8\_2}
\showDOI{\tempurl}


\bibitem[\protect\citeauthoryear{Fern{\'{a}}ndez, Mart{\'{\i}}nez{-}Prieto,
  Guti{\'{e}}rrez, Polleres, and Arias}{Fern{\'{a}}ndez et~al\mbox{.}}{2013}]%
        {fernandez2013binary}
\bibfield{author}{\bibinfo{person}{Javier~D. Fern{\'{a}}ndez},
  \bibinfo{person}{Miguel~A. Mart{\'{\i}}nez{-}Prieto},
  \bibinfo{person}{Claudio Guti{\'{e}}rrez}, \bibinfo{person}{Axel Polleres},
  {and} \bibinfo{person}{Mario Arias}.} \bibinfo{year}{2013}\natexlab{}.
\newblock \showarticletitle{Binary {RDF} representation for publication and
  exchange {(HDT)}}.
\newblock \bibinfo{journal}{\emph{J. Web Sem.}}  \bibinfo{volume}{19}
  (\bibinfo{year}{2013}), \bibinfo{pages}{22--41}.
\newblock
\urldef\tempurl%
\url{https://doi.org/10.1016/j.websem.2013.01.002}
\showDOI{\tempurl}


\bibitem[\protect\citeauthoryear{Fife}{Fife}{1968}]%
        {Fife68a}
\bibfield{author}{\bibinfo{person}{Dennis~W. Fife}.}
  \bibinfo{year}{1968}\natexlab{}.
\newblock \showarticletitle{{R68-47} Computer Scheduling Methods and Their
  Countermeasures}.
\newblock \bibinfo{journal}{\emph{{IEEE} Trans. Computers}}
  \bibinfo{volume}{17}, \bibinfo{number}{11} (\bibinfo{year}{1968}),
  \bibinfo{pages}{1098--1099}.
\newblock
\urldef\tempurl%
\url{https://doi.org/10.1109/TC.1968.226869}
\showDOI{\tempurl}


\bibitem[\protect\citeauthoryear{Garcia{-}Molina, Ullman, and
  Widom}{Garcia{-}Molina et~al\mbox{.}}{2009}]%
        {garcia2008database}
\bibfield{author}{\bibinfo{person}{Hector Garcia{-}Molina},
  \bibinfo{person}{Jeffrey~D. Ullman}, {and} \bibinfo{person}{Jennifer Widom}.}
  \bibinfo{year}{2009}\natexlab{}.
\newblock \bibinfo{booktitle}{\emph{Database systems - the complete book {(2.}
  ed.)}}.
\newblock \bibinfo{publisher}{Pearson Education}.
\newblock
\showISBNx{978-0-13-187325-4}


\bibitem[\protect\citeauthoryear{Graefe}{Graefe}{1993}]%
        {graefe1993query}
\bibfield{author}{\bibinfo{person}{Goetz Graefe}.}
  \bibinfo{year}{1993}\natexlab{}.
\newblock \showarticletitle{Query Evaluation Techniques for Large Databases}.
\newblock \bibinfo{journal}{\emph{{ACM} Comput. Surv.}} \bibinfo{volume}{25},
  \bibinfo{number}{2} (\bibinfo{year}{1993}), \bibinfo{pages}{73--170}.
\newblock
\urldef\tempurl%
\url{https://doi.org/10.1145/152610.152611}
\showDOI{\tempurl}


\bibitem[\protect\citeauthoryear{Graefe and McKenna}{Graefe and
  McKenna}{1993}]%
        {GraefeM93}
\bibfield{author}{\bibinfo{person}{Goetz Graefe} {and}
  \bibinfo{person}{William~J. McKenna}.} \bibinfo{year}{1993}\natexlab{}.
\newblock \showarticletitle{The Volcano Optimizer Generator: Extensibility and
  Efficient Search}. In \bibinfo{booktitle}{\emph{Proceedings of the Ninth
  International Conference on Data Engineering, April 19-23, 1993, Vienna,
  Austria}}. \bibinfo{publisher}{{IEEE} Computer Society},
  \bibinfo{pages}{209--218}.
\newblock
\urldef\tempurl%
\url{https://doi.org/10.1109/ICDE.1993.344061}
\showDOI{\tempurl}


\bibitem[\protect\citeauthoryear{Haas, Kossmann, Wimmers, and Yang}{Haas
  et~al\mbox{.}}{1997}]%
        {HaasKWY97Optimizing}
\bibfield{author}{\bibinfo{person}{Laura~M. Haas}, \bibinfo{person}{Donald
  Kossmann}, \bibinfo{person}{Edward~L. Wimmers}, {and} \bibinfo{person}{Jun
  Yang}.} \bibinfo{year}{1997}\natexlab{}.
\newblock \showarticletitle{Optimizing Queries Across Diverse Data Sources}. In
  \bibinfo{booktitle}{\emph{VLDB'97, Proceedings of 23rd International
  Conference on Very Large Data Bases, August 25-29, 1997, Athens, Greece}}.
  \bibinfo{publisher}{Morgan Kaufmann}, \bibinfo{pages}{276--285}.
\newblock


\bibitem[\protect\citeauthoryear{Hartig and Aranda}{Hartig and Aranda}{2016}]%
        {hartig2016bindings}
\bibfield{author}{\bibinfo{person}{Olaf Hartig} {and}
  \bibinfo{person}{Carlos~Buil Aranda}.} \bibinfo{year}{2016}\natexlab{}.
\newblock \showarticletitle{Bindings-Restricted Triple Pattern Fragments}. In
  \bibinfo{booktitle}{\emph{On the Move to Meaningful Internet Systems: {OTM}
  2016 Conferences - Confederated International Conferences: CoopIS, C{\&}TC,
  and {ODBASE} 2016, Rhodes, Greece, October 24-28, 2016, Proceedings}}
  \emph{(\bibinfo{series}{Lecture Notes in Computer Science})},
  Vol.~\bibinfo{volume}{10033}. \bibinfo{publisher}{Springer},
  \bibinfo{pages}{762--779}.
\newblock
\urldef\tempurl%
\url{https://doi.org/10.1007/978-3-319-48472-3\_48}
\showDOI{\tempurl}


\bibitem[\protect\citeauthoryear{Hartig, Bizer, and Freytag}{Hartig
  et~al\mbox{.}}{2009}]%
        {olaf2009}
\bibfield{author}{\bibinfo{person}{Olaf Hartig}, \bibinfo{person}{Christian
  Bizer}, {and} \bibinfo{person}{Johann~Christoph Freytag}.}
  \bibinfo{year}{2009}\natexlab{}.
\newblock \showarticletitle{Executing {SPARQL} Queries over the Web of Linked
  Data}. In \bibinfo{booktitle}{\emph{The Semantic Web - {ISWC} 2009, 8th
  International Semantic Web Conference, {ISWC} 2009, Chantilly, VA, USA,
  October 25-29, 2009. Proceedings}} \emph{(\bibinfo{series}{Lecture Notes in
  Computer Science})}, Vol.~\bibinfo{volume}{5823}.
  \bibinfo{publisher}{Springer}, \bibinfo{pages}{293--309}.
\newblock
\urldef\tempurl%
\url{https://doi.org/10.1007/978-3-642-04930-9\_19}
\showDOI{\tempurl}


\bibitem[\protect\citeauthoryear{Hartig, Letter, and P{\'{e}}rez}{Hartig
  et~al\mbox{.}}{2017}]%
        {olaf2017}
\bibfield{author}{\bibinfo{person}{Olaf Hartig}, \bibinfo{person}{Ian Letter},
  {and} \bibinfo{person}{Jorge P{\'{e}}rez}.} \bibinfo{year}{2017}\natexlab{}.
\newblock \showarticletitle{A Formal Framework for Comparing Linked Data
  Fragments}. In \bibinfo{booktitle}{\emph{The Semantic Web - {ISWC} 2017 -
  16th International Semantic Web Conference, Vienna, Austria, October 21-25,
  2017, Proceedings, Part {I}}} \emph{(\bibinfo{series}{Lecture Notes in
  Computer Science})}, Vol.~\bibinfo{volume}{10587}.
  \bibinfo{publisher}{Springer}, \bibinfo{pages}{364--382}.
\newblock
\urldef\tempurl%
\url{https://doi.org/10.1007/978-3-319-68288-4\_22}
\showDOI{\tempurl}


\bibitem[\protect\citeauthoryear{Heling, Acosta, Maleshkova, and
  Sure{-}Vetter}{Heling et~al\mbox{.}}{2018}]%
        {HelingAMS18}
\bibfield{author}{\bibinfo{person}{Lars Heling}, \bibinfo{person}{Maribel
  Acosta}, \bibinfo{person}{Maria Maleshkova}, {and} \bibinfo{person}{York
  Sure{-}Vetter}.} \bibinfo{year}{2018}\natexlab{}.
\newblock \showarticletitle{Querying Large Knowledge Graphs over Triple Pattern
  Fragments: An Empirical Study}. In \bibinfo{booktitle}{\emph{The Semantic Web
  - {ISWC} 2018 - 17th International Semantic Web Conference, Monterey, CA,
  USA, October 8-12, 2018, Proceedings, Part {II}}}. \bibinfo{pages}{86--102}.
\newblock
\urldef\tempurl%
\url{https://doi.org/10.1007/978-3-030-00668-6\_6}
\showDOI{\tempurl}


\bibitem[\protect\citeauthoryear{Kleinrock}{Kleinrock}{1964}]%
        {kleinrock1964analysis}
\bibfield{author}{\bibinfo{person}{Leonard Kleinrock}.}
  \bibinfo{year}{1964}\natexlab{}.
\newblock \showarticletitle{Analysis of A time-shared processor}.
\newblock \bibinfo{journal}{\emph{Naval research logistics quarterly}}
  \bibinfo{volume}{11}, \bibinfo{number}{1} (\bibinfo{year}{1964}),
  \bibinfo{pages}{59--73}.
\newblock


\bibitem[\protect\citeauthoryear{Neumann and Weikum}{Neumann and
  Weikum}{2010}]%
        {NeumannW10}
\bibfield{author}{\bibinfo{person}{Thomas Neumann} {and}
  \bibinfo{person}{Gerhard Weikum}.} \bibinfo{year}{2010}\natexlab{}.
\newblock \showarticletitle{The {RDF-3X} engine for scalable management of
  {RDF} data}.
\newblock \bibinfo{journal}{\emph{{VLDB} J.}} \bibinfo{volume}{19},
  \bibinfo{number}{1} (\bibinfo{year}{2010}), \bibinfo{pages}{91--113}.
\newblock
\urldef\tempurl%
\url{https://doi.org/10.1007/s00778-009-0165-y}
\showDOI{\tempurl}


\bibitem[\protect\citeauthoryear{P{\'{e}}rez, Arenas, and
  Guti{\'{e}}rrez}{P{\'{e}}rez et~al\mbox{.}}{2009}]%
        {perez2009semantics}
\bibfield{author}{\bibinfo{person}{Jorge P{\'{e}}rez}, \bibinfo{person}{Marcelo
  Arenas}, {and} \bibinfo{person}{Claudio Guti{\'{e}}rrez}.}
  \bibinfo{year}{2009}\natexlab{}.
\newblock \showarticletitle{Semantics and complexity of {SPARQL}}.
\newblock \bibinfo{journal}{\emph{{ACM} Trans. Database Syst.}}
  \bibinfo{volume}{34}, \bibinfo{number}{3} (\bibinfo{year}{2009}),
  \bibinfo{pages}{16:1--16:45}.
\newblock
\urldef\tempurl%
\url{https://doi.org/10.1145/1567274.1567278}
\showDOI{\tempurl}


\bibitem[\protect\citeauthoryear{Polleres, Kamdar, Fern{\'{a}}ndez, Tudorache,
  and Musen}{Polleres et~al\mbox{.}}{2018}]%
        {axeldesemweb18}
\bibfield{author}{\bibinfo{person}{Axel Polleres}, \bibinfo{person}{Maulik~R.
  Kamdar}, \bibinfo{person}{Javier~D. Fern{\'{a}}ndez}, \bibinfo{person}{Tania
  Tudorache}, {and} \bibinfo{person}{Mark~A. Musen}.}
  \bibinfo{year}{2018}\natexlab{}.
\newblock \showarticletitle{A More Decentralized Vision for Linked Data}. In
  \bibinfo{booktitle}{\emph{Proceedings of the 2nd Workshop on Decentralizing
  the Semantic Web co-located with the 17th International Semantic Web
  Conference, DeSemWeb@ISWC 2018, Monterey, California, USA, October 8, 2018.}}
\newblock


\bibitem[\protect\citeauthoryear{Saleem, Mehmood, and Ngomo}{Saleem
  et~al\mbox{.}}{2015}]%
        {SaleemMN15}
\bibfield{author}{\bibinfo{person}{Muhammad Saleem}, \bibinfo{person}{Qaiser
  Mehmood}, {and} \bibinfo{person}{Axel{-}Cyrille~Ngonga Ngomo}.}
  \bibinfo{year}{2015}\natexlab{}.
\newblock \showarticletitle{{FEASIBLE:} {A} Feature-Based {SPARQL} Benchmark
  Generation Framework}. In \bibinfo{booktitle}{\emph{The Semantic Web - {ISWC}
  2015 - 14th International Semantic Web Conference, Bethlehem, PA, USA,
  October 11-15, 2015, Proceedings, Part {I}}}. \bibinfo{pages}{52--69}.
\newblock
\urldef\tempurl%
\url{https://doi.org/10.1007/978-3-319-25007-6\_4}
\showDOI{\tempurl}


\bibitem[\protect\citeauthoryear{Schmachtenberg, Bizer, and
  Paulheim}{Schmachtenberg et~al\mbox{.}}{2014}]%
        {schmachtenberg2014adoption}
\bibfield{author}{\bibinfo{person}{Max Schmachtenberg},
  \bibinfo{person}{Christian Bizer}, {and} \bibinfo{person}{Heiko Paulheim}.}
  \bibinfo{year}{2014}\natexlab{}.
\newblock \showarticletitle{Adoption of the Linked Data Best Practices in
  Different Topical Domains}. In \bibinfo{booktitle}{\emph{The Semantic Web -
  {ISWC} 2014 - 13th International Semantic Web Conference, Riva del Garda,
  Italy, October 19-23, 2014. Proceedings, Part {I}}}
  \emph{(\bibinfo{series}{Lecture Notes in Computer Science})},
  Vol.~\bibinfo{volume}{8796}. \bibinfo{publisher}{Springer},
  \bibinfo{pages}{245--260}.
\newblock
\urldef\tempurl%
\url{https://doi.org/10.1007/978-3-319-11964-9\_16}
\showDOI{\tempurl}


\bibitem[\protect\citeauthoryear{Schmidt, Meier, and Lausen}{Schmidt
  et~al\mbox{.}}{2010}]%
        {Schmidt0L10}
\bibfield{author}{\bibinfo{person}{Michael Schmidt}, \bibinfo{person}{Michael
  Meier}, {and} \bibinfo{person}{Georg Lausen}.}
  \bibinfo{year}{2010}\natexlab{}.
\newblock \showarticletitle{Foundations of {SPARQL} query optimization}. In
  \bibinfo{booktitle}{\emph{Database Theory - {ICDT} 2010, 13th International
  Conference, Lausanne, Switzerland, March 23-25, 2010, Proceedings}}.
  \bibinfo{publisher}{{ACM}}, \bibinfo{pages}{4--33}.
\newblock
\urldef\tempurl%
\url{https://doi.org/10.1145/1804669.1804675}
\showDOI{\tempurl}


\bibitem[\protect\citeauthoryear{Schwarte, Haase, Hose, Schenkel, and
  Schmidt}{Schwarte et~al\mbox{.}}{2011}]%
        {schwarte2011fedx}
\bibfield{author}{\bibinfo{person}{Andreas Schwarte}, \bibinfo{person}{Peter
  Haase}, \bibinfo{person}{Katja Hose}, \bibinfo{person}{Ralf Schenkel}, {and}
  \bibinfo{person}{Michael Schmidt}.} \bibinfo{year}{2011}\natexlab{}.
\newblock \showarticletitle{FedX: Optimization Techniques for Federated Query
  Processing on Linked Data}. In \bibinfo{booktitle}{\emph{The Semantic Web -
  {ISWC} 2011 - 10th International Semantic Web Conference, Bonn, Germany,
  October 23-27, 2011, Proceedings, Part {I}}} \emph{(\bibinfo{series}{Lecture
  Notes in Computer Science})}, Vol.~\bibinfo{volume}{7031}.
  \bibinfo{publisher}{Springer}, \bibinfo{pages}{601--616}.
\newblock
\urldef\tempurl%
\url{https://doi.org/10.1007/978-3-642-25073-6\_38}
\showDOI{\tempurl}


\bibitem[\protect\citeauthoryear{Verborgh, Sande, Hartig, Herwegen, Vocht,
  Meester, Haesendonck, and Colpaert}{Verborgh et~al\mbox{.}}{2016}]%
        {verborgh2016triple}
\bibfield{author}{\bibinfo{person}{Ruben Verborgh},
  \bibinfo{person}{Miel~Vander Sande}, \bibinfo{person}{Olaf Hartig},
  \bibinfo{person}{Joachim~Van Herwegen}, \bibinfo{person}{Laurens~De Vocht},
  \bibinfo{person}{Ben~De Meester}, \bibinfo{person}{Gerald Haesendonck}, {and}
  \bibinfo{person}{Pieter Colpaert}.} \bibinfo{year}{2016}\natexlab{}.
\newblock \showarticletitle{Triple Pattern Fragments: {A} low-cost knowledge
  graph interface for the Web}.
\newblock \bibinfo{journal}{\emph{J. Web Sem.}}  \bibinfo{volume}{37-38}
  (\bibinfo{year}{2016}), \bibinfo{pages}{184--206}.
\newblock
\urldef\tempurl%
\url{https://doi.org/10.1016/j.websem.2016.03.003}
\showDOI{\tempurl}


\bibitem[\protect\citeauthoryear{Weiss, Karras, and Bernstein}{Weiss
  et~al\mbox{.}}{2008}]%
        {WeissKB08}
\bibfield{author}{\bibinfo{person}{Cathrin Weiss}, \bibinfo{person}{Panagiotis
  Karras}, {and} \bibinfo{person}{Abraham Bernstein}.}
  \bibinfo{year}{2008}\natexlab{}.
\newblock \showarticletitle{Hexastore: sextuple indexing for semantic web data
  management}.
\newblock \bibinfo{journal}{\emph{{PVLDB}}} \bibinfo{volume}{1},
  \bibinfo{number}{1} (\bibinfo{year}{2008}), \bibinfo{pages}{1008--1019}.
\newblock
\urldef\tempurl%
\url{https://doi.org/10.14778/1453856.1453965}
\showDOI{\tempurl}


\bibitem[\protect\citeauthoryear{Wilschut and Apers}{Wilschut and
  Apers}{1993}]%
        {wilschut1993dataflow}
\bibfield{author}{\bibinfo{person}{Annita~N. Wilschut} {and}
  \bibinfo{person}{Peter M.~G. Apers}.} \bibinfo{year}{1993}\natexlab{}.
\newblock \showarticletitle{Dataflow Query Execution in a Parallel Main-memory
  Environment}.
\newblock \bibinfo{journal}{\emph{Distributed and Parallel Databases}}
  \bibinfo{volume}{1}, \bibinfo{number}{1} (\bibinfo{year}{1993}),
  \bibinfo{pages}{103--128}.
\newblock
\urldef\tempurl%
\url{https://doi.org/10.1007/BF01277522}
\showDOI{\tempurl}


\end{thebibliography}

\end{document}